\documentclass[10pt,aps,floatfix,prd,twocolumn,merge,nofootinbib,
        superscriptaddress,notitlepage,longbibliography]{revtex4} 
\usepackage{graphicx} 
\usepackage{dcolumn} 
\usepackage{amsmath} 
\usepackage{slashed}
\usepackage{cancel}
\usepackage[caption=false]{subfig}
\usepackage[usenames,dvipsnames,svgnames]{xcolor}
\usepackage{braket}
\usepackage{multirow}
\usepackage[shortlabels]{enumitem}

\newcommand{\be}{\begin{equation}}
\newcommand{\ee}{\end{equation}}
\newcommand{\bea}{\begin{eqnarray}}
\newcommand{\eea}{\end{eqnarray}}

\usepackage{hyperref} 

\usepackage[capitalise]{cleveref}

\hypersetup{%
  colorlinks=true, linktocpage=true, pdfstartpage=1, pdfstartview=FitV,%
  breaklinks=true, pdfpagemode=UseNone, pageanchor=true, pdfpagemode=UseOutlines,%
  plainpages=false, bookmarksnumbered, bookmarksopen=true, bookmarksopenlevel=1,%
  hypertexnames=true, pdfhighlight=/O,
  urlcolor=Cerulean, linkcolor=Cerulean, citecolor=Cerulean, 
}

\usepackage[normalem]{ulem}

\begin{document}
\title{Probing Multicomponent Extension of Inert Doublet Model with a Vector Dark Matter}
\author{Sreemanti Chakraborti}
\email{sreemanti@iitg.ac.in}
\affiliation{Indian Institute of Technology Guwahati, Guwahati 781 039, India}
\author{Amit Dutta Banik}
\email{amitdbanik@iitg.ac.in}
\affiliation{Indian Institute of Technology Guwahati, Guwahati 781 039, India}
\author{Rashidul Islam}
\email{rislam@iitg.ac.in}
\affiliation{Indian Institute of Technology Guwahati, Guwahati 781 039, India}
%

%
\begin{abstract}
Although theoretically well-motivated as a minimal consistent Dark Matter (DM) model, the Inert Doublet Model (IDM) fell short in explaining the existence of DM in the intermediate mass region ($100-500$~GeV). We try to address this problem by introducing an additional vector dark matter. We show that the relic density of inert dark matter candidate can be enhanced significantly with new interactions in the intermediate mass region $100-500$~GeV in the multicomponent dark matter model when compared with the usual single component inert doublet model. We also show that we can get a reasonable excess in the $\slashed{E}_T$ distribution if we do not apply a very hard $\slashed{E}_T$ cut on it as is customary in any dark matter search at the LHC.
\end{abstract}

\maketitle

%

%
\section{Introduction}
\label{sec:intro}
The observational results from the satellite-borne experiment WMAP~\cite{Hinshaw:2012aka} and more recently Planck~\cite{Aghanim:2018eyx} have now firmly established the presence of dark matter in the Universe. Their results reveal that more than 80\% of matter content of the Universe is in the form of mysterious unknown matter called the dark matter (DM). Until now, only the gravitational interactions of DM have been manifested by most of its indirect evidence namely the flatness of rotation curves of spiral galaxies, gravitational lensing, phenomena of Bullet cluster and other various colliding galaxy clusters etc. However, the particle nature of DM still remains an enigma. There are various ongoing dark matter direct detection experiments such as LUX~\cite{Akerib:2016vxi}, XENON1T~\cite{Aprile:2018dbl}, PandaX-II~\cite{Cui:2017nnn} and indirect detection experiments like Fermi-LAT~\cite{Fermi-LAT:2016uux} and H.E.S.S~\cite{Abdallah:2018qtu}, which have been trying to investigate the particle nature as well as the interaction type (spin-dependent or spin-independent) of DM with the visible sector by measuring the recoil energy of the scattered detector nuclei. However, the null results of these experiments have severely constrained the DM-nucleon spin-independent scattering cross section $\sigma_{\rm SI}$. The overwhelming success of the Standard Model (SM) has been established undoubtedly by the discovery of Higgs by ATLAS~\cite{Aad:2012tfa} and CMS~\cite{Chatrchyan:2012xdj}. However, the SM of particle physics is inadequate to explain the riddles of neutrino mass, dark matter, inflation etc. In the Standard Model (SM) there is no dark matter candidate and one should, therefore, look beyond. Depending upon the production mechanism in the early Universe, the dark matter can be called thermal or non-thermal. In the former case, dark matter particles were in both thermal as well as in chemical equilibrium with other particles in the thermal soup at a very early epoch. As the Universe expands and temperature decreases, the thermal dark matter candidate freezes out and becomes a relic. Weakly Interacting Massive Particle (WIMP)~\cite{Griest:2000kj,Bertone:2004pz} is the most studied candidate for the thermal dark matter scenario. WIMP candidates are also referred to as cold dark matter (CDM) for being non-relativistic at the time of decoupling from the thermal bath. Direct search experiments mainly search for WIMP-like DM candidates.

Various WIMP dark matter scenarios have been proposed and explored extensively in the literature. Among these extensions of the SM, a simple approach is to introduce an additional $SU(2)$ scalar doublet which produces no vacuum expectation value (\emph{vev}) due to the imposed $Z_2$ symmetry under which the doublet is odd. The resulting model is named the Inert Doublet Model (IDM). Dark matter phenomenology of IDM has been elaborately studied in the literatures~\cite{Ma:2006km,Barbieri:2006dq,Cirelli:2005uq,LopezHonorez:2006gr,Gustafsson:2007pc,Cao:2007rm,Majumdar:2006nt,Lundstrom:2008ai,Honorez:2010re,LopezHonorez:2010tb,Chowdhury:2011ga,Borah:2012pu,Arhrib:2013ela,Chakrabarty:2015yia,Plascencia:2015xwa,Borah:2017dfn,Ahriche:2017iar,Borah:2018rca}. Apart from the dark matter phenomenology, collider searches from IDM are also performed extensively~\cite{Miao:2010rg,Gustafsson:2012aj,Hashemi:2016wup,Datta:2016nfz,Poulose:2016lvz,Goudelis:2013uca,Arhrib:2013ela,Alves:2016bib,Aoki:2013lhm,Kanemura:2016sos}. IDM can provide a low mass dark matter candidate with mass smaller than 130~GeV and a high mass regime above 500~GeV~\cite{Plascencia:2015xwa,Ahriche:2017iar}. In the region in between, the inert dark matter is found to be underabundant due to the large annihilation into gauge bosons. However, this can be resolved if one assumes that dark matter is multicomponent in nature. Different multicomponent dark matter models including both thermal and non-thermal dark matter candidates have been explored in the literatures~\cite{Biswas:2013nn,Bhattacharya:2013hva,Bian:2013wna,Modak:2013jya,Bian:2014cja,Biswas:2014hoa,Biswas:2015sva,Bhattacharya:2016ysw,Arcadi:2016kmk,Aoki:2016glu,DuttaBanik:2016jzv,Pandey:2017quk,Ahmed:2017dbb,Zakeri:2018hhe,Chakraborti:2018lso,YaserAyazi:2018lrv,Bhattacharya:2018cgx,Chialva:2012rq,Karam:2016rsz,Herrero-Garcia:2018qnz}. In multicomponent dark matter models one of the candidates shares a fraction of total DM relic abundance and the other candidate provides the required amount of relic density in order to satisfy the total DM relic density observed by Planck~\cite{Aghanim:2018eyx}. In this work, we consider another vector boson dark matter candidate in addition to the existing IDM. While the IDM is odd under a discrete $Z_2$ symmetry, the added vector boson is also considered to be odd under another $Z_2'$ symmetry making both the candidates stable. There are other possible ways to stabilise the vector such as the inclusion of a dark $U(1)_D$ symmetry~\cite{Ko:2016ala,Ahmed:2017dbb}. Such scenarios include an extra scalar field that mixes with the SM Higgs boson after symmetry breaking and provides a stable vector boson dark matter. Here, we consider a minimal scenario avoiding such mixings and stable vector DM is achieved only by introducing a \(Z_2\) symmetry. In fact, we have observed that in certain conditions it is even possible to boost the relic density contribution from inert doublet dark matter. This multicomponent scenario, therefore, provides a window to explore the inert dark matter in the mass range 130-500~GeV. In addition, the inclusion of a vector dark matter candidate enriches the dark sector. In this work, we present the phenomenology of the proposed multicomponent dark matter model and test the viability of the model by constraining the model parameter space from different theoretical and experimental bounds. We investigate the possibility whether the relic density of inert dark matter candidate can be enhanced significantly in the intermediate regime of inert doublet mass $\sim$130-500~GeV. The proposed multicomponent dark matter model further opens up a new window of IDM to be tested in the LHC for possible signatures and promising outcomes. 

The paper is organised as follows. In \cref{sec:model} we describe our model including its field contents and Lagrangian. \cref{sec:bounds} gives the theoretical and experimental bounds on the various model parameters. The formalism which includes the form of the Boltzmann equations for the relic density and expressions for the direct detection bounds on the dark matter is described in \cref{sec:formalism}. We show in \cref{sec:results} the detail observations and results from the dark matter analysis and in \cref{sec:collider}, the outcome of the collider searches in the present LHC environment. Finally, we conclude our paper in \cref{sec:conc}.

%
\section{The Model}
\label{sec:model}
In the present work, we consider a multicomponent dark matter model by including an additional vector boson to the Inert Doublet Model (IDM) of dark matter. The inert doublet is considered to be odd under a $Z_2$ which ensures that it does not acquire any vacuum expectation value (\emph{vev}) after spontaneous symmetry breaking. Also, particles in the Standard Model (SM) are even under this $Z_2$ which forbids the decay of the lightest inert particle. Therefore, the lightest inert particle can serve as a dark matter (DM) candidate. Similarly, we impose another discrete $Z_2'$ symmetry upon the vector boson $X$ which can also be a feasible DM candidate in our model. We denote the SM Higgs doublet as $H$ while the inert Higgs doublet is $\Phi$. The total Lagrangian of our study is as follows
\begin{multline}
 {\cal L} = |D_\mu H|^2 + |D_\mu \Phi|^2 - V_{\rm IDM}
          \\
          + {\cal L}_{H/\rm \Phi, X} + {\cal L}_{\rm X}
          + {\cal L}_{\rm Yuk} + {\cal L}^{\cancel{\rm Higgs}}_{\rm SM} ,
 \label{Lag_tot}
\end{multline}
where ${\cal L}^{\cancel{\rm Higgs}}_{\rm SM}$ is the SM Lagrangian sans the Higgs part, ${\cal L}_{\rm Yuk}$ is the SM Yukawa interaction Lagrangian. ${\cal L}_{\rm X}$ is the vector DM sector which is given by
\begin{gather}
  {\cal L}_{\rm X}
   = - \frac{1}{4} X^{\mu\nu} X_{\mu\nu}
     + \frac{1}{2} M^2_X X_{\mu}X^{\mu} + \frac{1}{4}\lambda_X (X_{\mu}X^{\mu})^2\,,
\end{gather}
the interaction term, ${\cal L}_{H/\rm \Phi, X}$ between the scalar fields $H,\Phi$ and the dark vector boson, $X$ is
\begin{gather}
  {\cal L}_{H/\rm \Phi, X}
   = \lambda_{HX} (X^\mu X_\mu) (H^\dag H)
   + \lambda_{\Phi X} (X^\mu X_\mu) (\Phi^\dag \Phi)\,;
\end{gather}
and finally, the well known IDM potential
\begin{multline}
 V_{\rm IDM}
   = m_{11}^2 (H^\dag H) + m_{22}^2 (\Phi^\dag\Phi)
     \\
     + \lambda_1 (H^\dag H)^2 + \lambda_2 (\Phi^\dag\Phi)^2
     + \lambda_3(H^\dag H)(\Phi^\dag\Phi)
     \\
     + \lambda_4(H^\dag\Phi)(\Phi^\dag H)
     + \frac{\lambda_5}{2}\left[(H^\dag\Phi)^2 + {\rm h.c} \right]\, ,
 \label{pot_IDM}
\end{multline}
where all the couplings $\lambda_i, i=1-5$ are assumed to be real and also $m_{11}^2<0$ so that spontaneous symmetry breaking (SSB) occurs. After SSB the SM Higgs doublet receives a \emph{vev} $v=246$~GeV and the doublet fields are denoted as
\begin{align}
 H = \begin{pmatrix}
      0 \\
      \frac{1}{\sqrt{2}}(v+h) 
     \end{pmatrix} \, ,
 \qquad
 \Phi = \begin{pmatrix}
         H^+ \\
         \frac{1}{\sqrt{2}}(H_0+iA_0) 
        \end{pmatrix} \, .
 \label{fields}
\end{align}
Masses of different physical scalar including SM Higgs and inert particles and the vector boson $X$ of the dark sector are given as follows
\begin{align}
\begin{aligned}
 m_h^2 =\,& 2 \lambda_1 v^2
 \\
 m_{H^{\pm}}^{2} =\,& m_{22}^{2}+\lambda_{3}\frac{v^{2}}{2}
 \\
 m_{H_0}^{2} =\,& m_{22}^{2}+(\lambda_{3}+\lambda_{4}+\lambda_{5})\frac{v^{2}}{2}
 \\
 m_{A_0}^{2} =\,& m_{22}^{2}+(\lambda_{3}+\lambda_{4}-\lambda_{5})\frac{v^{2}}{2}
 \\
 m_{X}^{2} =\,& M^2_X + \lambda_{HX}{v^2} \,.
\end{aligned}
\label{mass}
\end{align}
In the above \cref{mass}, $m_h$ refers to mass of SM Higgs boson ($h$) $m_h=125.09$~GeV~\cite{Patrignani:2016xqp}. As mentioned, even after SSB the dark $Z_2$ ($Z_2'$) symmetry of inert doublet (vector boson $X$) remains intact and the lightest scalar $H_0$ of the IDM\footnote{We consider $\lambda_5<0$.} along with $X$ provides a scenario for multicomponent dark matter.

Before we present the discussions on the phenomenology of multicomponent dark matter model, we first mention some important theoretical and experimental bounds to be taken into account in the next section.

%
\section{Constraints and bounds}
\label{sec:bounds}
\begin{description}[leftmargin=0pt,labelindent=0pt]
 \item[{\bf Vacuum Stability:}] In order to stabilise the Higgs vacuum following conditions must be satisfied~\cite{Kannike:2012pe}
 \begin{gather}
   \begin{gathered}
     \lambda_1,\,\lambda_2 > 0\, ,
     \lambda_3 + 2\sqrt{\lambda_1\lambda_2} > 0\, ,
     \\
     \lambda_3 + \lambda_4 - |\lambda_5| + 2\sqrt{\lambda_1\lambda_2} > 0\, .
   \end{gathered}
   \label{4}
 \end{gather}

 \item[{\bf Perturbativity:}] Quartic interactions at tree level at high energy provides bound on the eigenvalues $|\Lambda_i|$ of quartic couplings which must obey the relation $|\Lambda_i|\leq 4\pi$.

 \item[{\bf LEP:}] LEP~\cite{Patrignani:2016xqp} provides bound from the decay width measurement of 
$Z$ boson which is given as
 \begin{gather}
   \begin{gathered}
     m_{H_0} + m_{A_0} > m_Z \, , \\
     m_{H^{\pm}} > 80 ~\rm{GeV}.
   \end{gathered}
 \end{gather}

 \item[{\bf Relic Density:}] In the present model we have two feasible dark matter candidates $H_0$ and $X$. Since both the candidates contribute to the dark matter relic density obtained from Planck~\cite{Aghanim:2018eyx} experiment, one must satisfy the following relation
 \begin{gather}
   \Omega_{\rm DM} {\rm h}^2 = 0.1199{\pm 0.0027}= \Omega_{H_0} {\rm h}^2 + 
   \Omega_{X} {\rm h}^2\, ,
 \label{planck}
 \end{gather}
where ${\rm h}$ denotes the Hubble parameter (100 km~s$^{-1}$~Mpc$^{-1}$) and relic density of inert doublet (vector dark matter) is given as $\Omega_{H_0}{\rm h}^2$ ($\Omega_X{\rm h}^2$).

 \item[{\bf Direct Detection Experiments:}] Apart from satisfying the conditions from dark matter relic density, both the dark matter candidates must be in agreement with present direct detection bounds from different dark matter search experiments as well. In this work, we constrain the model parameter space with the latest direct detection limits obtained from XENON1T~\cite{Aprile:2018dbl}.

 \item[{\bf Indirect Detection Experiments:}] Dark matter is further constrained from the observations of diffused $\gamma$-rays from the Galactic Centre (GC) and dwarf spheroidal galaxies (dSphs) where DM density appears to be large. Experiments such as Fermi-LAT~\cite{Fermi-LAT:2016uux} and H.E.S.S~\cite{Abdallah:2018qtu} has put constraints on the upper limit of velocity averaged scattering cross sections for various channels which can contribute to the observed photon flux. Here we obeyed constraints on the model parameter space emanating from the observations of the latest Fermi-LAT data~\cite{Fermi-LAT:2016uux}.

 \item[{\bf Searches at LHC:}] The DM searches at the LHC has been performed in various hadronic as well as leptonic channels. Here we will follow the \emph{dilepton + missing energy} ($2\ell + \slashed{E}_T$) searches for our present study. Usually, such experimental searches were conducted in the context of Supersymmetric (SUSY) theories. Since our study does not include SUSY, we will follow the outcome of those searches with some care.
\end{description}

%
\section{Formalism for Dark Matter Analysis}
\label{sec:formalism}
Before we begin our analysis of the multicomponent dark matter scenario, we briefly mention the calculations of relic density and direct detection measurements in the present model.

\subsection{Relic density of dark matter candidates}
Since the present model deals with two dark matter candidates which also interact with themselves, we have to solve for the coupled Boltzmann equation. Relic density for each of the dark matter candidate is obtained by solving these coupled equations which are written as
\begin{multline}
  \frac{{\rm d} n_{H_0}}{{\rm d} t} + 3 {\rm H}n_{H_0}
  =
  - \langle \sigma {\rm v}\rangle_{H_0H_0\to SM SM} \,
    \left( n_{H_0}^{2} - n_{H_0\rm{eq}}^{2} \right)
  \\
  + \langle \sigma {\rm v}\rangle_{X X \to H_0H_0}
    \left( n_{X}^{2} - \frac{n_{X\rm{eq}}^2}{n_{H_0\rm{eq}}^{2}}n_{H_0}^2 \right)\, ;
  \label{Boltzmann1}
\end{multline}
\begin{multline}
  \frac{{\rm d} n_{X}}{{\rm d} t} + 3 {\rm H}n_{X}
  =
  - \langle \sigma {\rm v}\rangle_{XX\to SM SM} \left( n_{X}^{2} - n_{X\rm{eq}}^{2} \right)
  \\
  - \langle \sigma {\rm v}\rangle_{X X \to H_0H_0}
  \left( n_{X}^{2} - \frac{n_{X\rm{eq}}^2}{n_{H_0\rm{eq}}^{2}}n_{H_0}^2 \right)\, .
  \label{Boltzmann2}
\end{multline}
where ${\rm m_X} > {\rm m_{H_0}}$. Similarly, for ${\rm m_{H_0}} > {\rm m_X}$ ,
\begin{multline}
  \frac{{\rm d} n_{H_0}}{{\rm d} t} + 3 {\rm H}n_{H_0}
  =
  - \langle \sigma {\rm v}\rangle_{H_0H_0\to SM SM}
    \left( n_{H_0}^{2} - n_{H_0\rm{eq}}^{2} \right)
  \\
  - \langle \sigma {\rm v}\rangle_{H_0 H_0 \to XX}
    \left( n_{H_0}^{2} - \frac{n_{H_0\rm{eq}}^2}{n_{X\rm{eq}}^{2}}n_{X}^2 \right)\, ;
  \label{Boltzmann3}
\end{multline}
\begin{multline}
  \frac{{\rm d} n_{X}}{{\rm d} t} + 3 {\rm H}n_{X}
  =
  - \langle \sigma {\rm v}\rangle_{XX\to SM SM}
    \left( n_{X}^{2} - n_{X\rm{eq}}^{2} \right)
  \\
  + \langle \sigma {\rm v}\rangle_{H_0 H_0 \to X X}
    \left( n_{H_0}^{2} - \frac{n_{H_0\rm{eq}}^2}{n_{X\rm{eq}}^{2}}n_{X}^2 \right)\, .
  \label{Boltzmann4}
\end{multline}
In the above \cref{Boltzmann1,Boltzmann2}, $n_{i}, i=X,H_0$, denotes the number density of dark matter particles and their equilibrium number densities are expressed as $n_{i\rm{eq}}$ respectively. Annihilation cross section of dark matter candidates into SM are given as $\langle \sigma {\rm v}\rangle_{ii\to SMSM}$ while the same between themselves are denoted by the $\langle \sigma {\rm v}\rangle_{X X \to H_0H_0}$  ($\langle \sigma {\rm v}\rangle_{H_0 H_0 \to X X}$) for $m_X>m_{H_0}$ ($m_{H_0}>m_X$). Solving for the \cref{Boltzmann1,Boltzmann2} (or \cref{Boltzmann3,Boltzmann4} depending on the masses of $m_{H_0}$ and $m_{X}$), one can obtain the relic density contributions from each of the dark matter candidates of the form
\begin{gather}
  \Omega_i {\rm h}^2=2.755\times10^8\frac{m_i}{GeV} \, Y_i(T_0) \, ,
   \quad i=H_0,X\, ,
  \label{relic}
\end{gather}
where $Y_i=n_i/s$, is the yield of dark matter candidate obtained at present temperature of Universe $T_0$ and $s$ is the entropy density of the Universe. The total DM relic density is then obtained by adding individual relic density of both the candidates as mentioned in \cref{planck}. It is to be noted that the annihilation cross section $\langle \sigma {\rm v}\rangle_{X X \to H_0H_0}$ ($\langle \sigma {\rm v}\rangle_{H_0 H_0 \to X X}$ ) depends on the coupling between dark sector particles and therefore the coupling $\lambda_{\Phi X}$ plays a significant role in the dark matter phenomenology.

\subsection{Direct detection of DM candidates}
Dark matter direct search experiments like LUX, XENON1T etc. search for direct interactions of dark matter with detector nuclei. Dark matter candidate can undergo elastic scattering with detector nuclei and recoil energy will be transferred which can be measured at the detector. However, no such event has been recorded yet which provides a stringent limit on dark matter direct detection cross section. In the present model, both of dark matter candidates $H_0$ and $X$ can undergo spin-independent elastic scattering with the detector nuclei. Since, the model involves two dark matter candidates, the final direct detection cross section for each dark matter will be scaled by a factor $r_i =\Omega_i {\rm h}^2 / \Omega_{DM}{\rm h}^2, i=H_0,X$. Bounds from direct detection of dark matter will constrain the model parameters. It is to be noted that coupling $\lambda_{\Phi X}$ has no contribution in direct detection measurements. The spin independent direct detection cross section for the scalar dark matter $H_0$ is given as
\begin{gather}
  \sigma_{\rm {H_0}}^{\rm {SI}}=r_{H_0} \frac{\lambda_L^2}{16\pi}\frac{1}{m_h^4} f^2
  \frac{m_N^4}{(m_{H_0}+m_N)^2},
  \label{scalardd}
\end{gather}
where $\lambda_L = ( \lambda_3 + \lambda_4 + \lambda_5 ) / 2$ and $m_N$ denotes the mass of the nucleon. In the above expression of \cref{scalardd} the contributions from nuclear matrix elements are given by the factor $f\sim0.3$~\cite{Barbieri:2006dq,Alarcon:2011zs}. Similarly, the spin-independent direct detection cross section for the vector dark matter candidate $X$ is expressed as
\begin{gather}
  \sigma_{\rm {X}}^{\rm {SI}}
  =
  r_{X} \frac{\lambda_{HX}^2}{16\pi}\frac{1}{m_h^4} f^2 \frac{m_N^4}{(m_{X}+m_N)^2} \, ,
  \label{vectordd}
\end{gather}

We constrain the model parameter space using the most stringent direct detection bounds obtained from XENON1T~\cite{Aprile:2018dbl}.

%
\section{Observations and results}
\label{sec:results}
Before we present the discussions on our model with multicomponent dark matter, we summarise the main parameters in the model. The model is implemented in  FeynRules~\cite{Alloul:2013bka} and the relic density computation is performed using micrOMEGAS~\cite{Belanger:2014vza} package scanning over the available parameter space. As mentioned previously, the model has two dark matter candidates, the lightest inert doublet particle $H_0$ and the vector boson $X$. The parameters from the inert doublet that contribute to relic density and direct detection measurements are well known
\[
  \lambda_L,~m_{H_0},~m_{A_0},~m_{H^\pm}.
\]
Similarly, the parameters that contribute to DM phenomenology of the vector boson are
\[
  \lambda_{HX},~m_{X}.
\]
Apart from these above-mentioned parameters, there is another coupling $\lambda_{{\rm \Phi} X}$ which can contribute to the annihilation among the dark sector particles. It is to be noted that even for the coupling $\lambda_{{\rm \Phi}X}=0$, the above annihilation can occur through Higgs mediated diagrams. However, such contributions are significant only near the Higgs resonance and become small when we consider the mass of dark matter candidates away from Higgs resonance. Also, the strong bound from dark matter direct detection severely constrains the couplings $\lambda_L$ and $\lambda_{HX}$ reducing the effects of dark sector annihilation. On the other hand, the four-point coupling $\lambda_{{\rm \Phi}X}$ is a completely independent parameter which does not contribute to dark matter direct detection but can contribute to relic density of DM particles. In this work, our primary aim is to study how the $2\leftrightarrow 2$ annihilation between dark sector particles $H_0$ and $X$ affects the multicomponent dark matter scenario. To this end, we consider the intermediate regimes of dark matter masses in our work
\begin{gather}
  \boxed{100~{\rm~GeV}\leq m_{H_0},m_{X}\leq 500~{\rm~GeV}\, .}
  \label{mass_range}
\end{gather} 
It is to be noted that for the case of the inert doublet, co-annihilation effects can be significant if mass splitting between the scalar $H_0$, $A_0$ and the charged particle $H^\pm$ is small. In fact, for a pure inert doublet dark matter, relic density is negligible for large mass splitting. In this work, we present our results for two values of mass splitting $\Delta m=10$~GeV and $25$~GeV where $\Delta m = m_{A_0} - m_{H_0}$ (in~GeV) and $m_{H^\pm} = m_{A_0} + 0.1$~GeV\footnote{$m_{H^\pm}-m_{A_0}=0.1$~GeV is maintained throughout the analysis for all the different cases considered in the work.}. The coupling $\lambda_L$ and $\lambda_{HX}$ should not be large in order to satisfy direct detection bounds and we set them to be equal, $\lambda_L = \lambda_{HX} = 0.01$. In this way, we restrict other model parameters and use the coupling $\lambda_{\Phi X}$ as a variable, the new parameter to determine the allowed regions in this framework.

To begin with, we first consider a simplified case setting $\lambda_L=\lambda_{HX}=0$ and changing $\lambda_{\Phi X}$ from a very small value to $0.1$\footnote{It is to be noted that since we have already set the coupling $\lambda_{HX}=0$, we cannot use $\lambda_{\Phi X}=0$ when solving the coupled Boltzmann equation. Hence we considered a very small $\lambda_{\Phi X}\sim 10^{-15}$ such that it reproduces the nature of IDM.}. We denote the relic density of inert dark matter $H_0$ as $\Omega_{H_0}{\rm h}^2$ and that of the vector DM candidate as $\Omega_X{\rm h}^2$ (as expressed in \cref{planck}).
\begin{figure}[!ht]
  \centering
  \includegraphics[width=\linewidth]{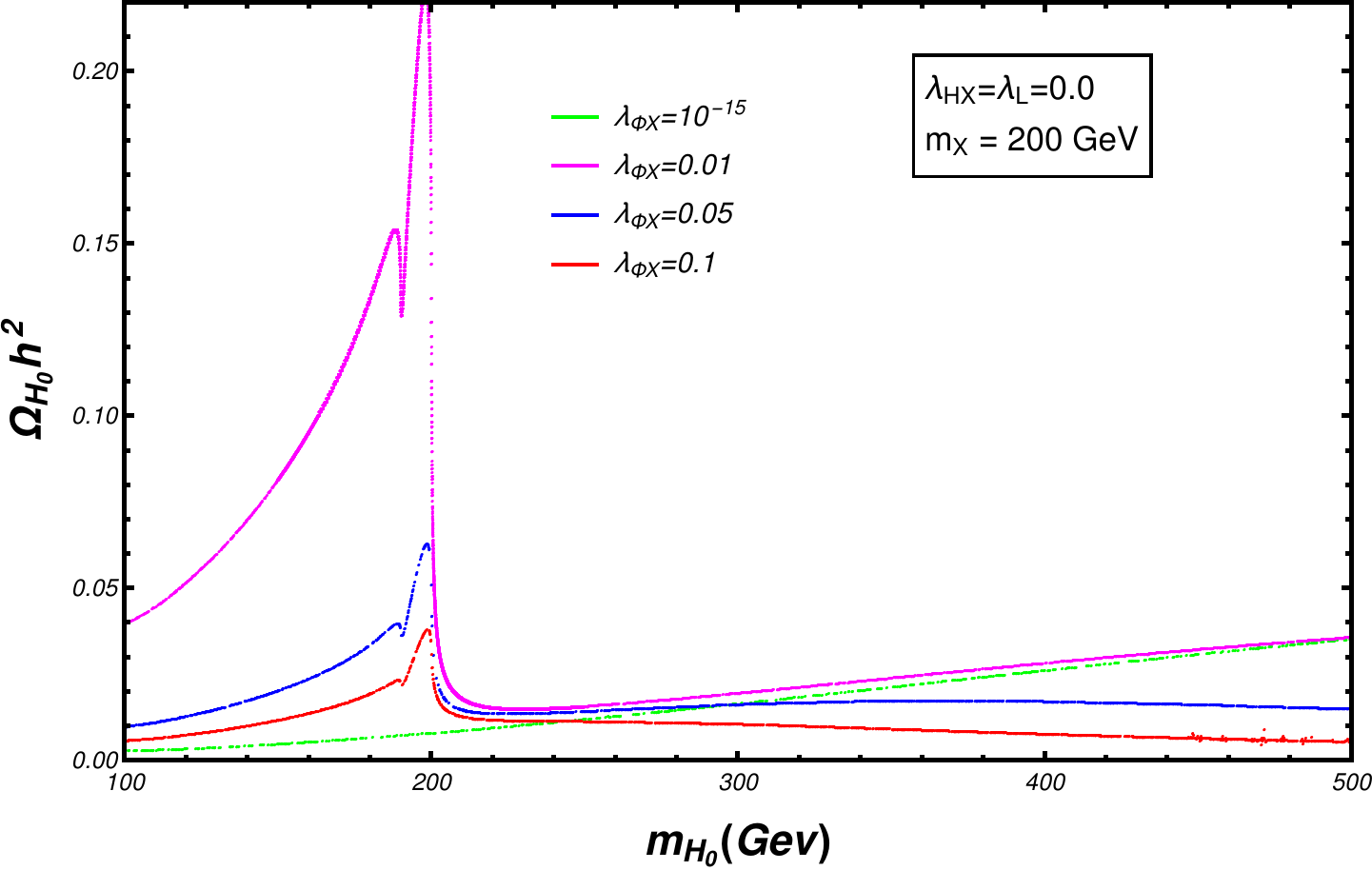}
  \caption{$\Omega_{H_0}{\rm h}^2$ vs $m_{H_0}$ for $\Delta$m$=10$~GeV. Only the quartic interactions in $X\ X\to H_0H_0$ and quartic gauge interactions in $H_0H_0\to SM\ SM$ are considered.}
  \label{fig:fig1}
\end{figure}
In \cref{fig:fig1} we show the variation of $\Omega_{H_0}{\rm h}^2$ against $m_{H_0}$ for the chosen values of couplings $\lambda_{\Phi X}$ for $\lambda_{L}=\lambda_{HX}=0$. With this consideration, only the gauge interactions of inert dark matter will survive and there will be $H_0H_0\leftrightarrow XX$ annihilation due to quartic coupling $\lambda_{\Phi X}$ only. Moreover, with $\lambda_L=\lambda_{HX}=0$, there will not be any direct detection signatures of both the dark matter candidates. We set the mass difference $\Delta m=m_A-m_{H_0}=10$~GeV and $m_X=200$~GeV. From \cref{fig:fig1}, we observe that for $\lambda_{\Phi X}=0.01$, the relic density $\Omega_{H_0}{\rm h}^2$ of $H_0$ changes drastically from the case with $\lambda_{\Phi X}=10^{-15}$, where $H_0$ can annihilate only to SM gauge bosons for the region of mass $m_{H_0}\leq 200$~GeV and afterwards follows the same pattern of IDM ($\lambda_{\Phi X}=10^{-15}$). The increase in $H_0$ relic density happens till $m_{H_0}\sim m_X$ and it falls sharply as $m_{H_0}\geq m_X$ when $H_0H_0\to XX$ annihilation channel opens. Apart from that there exists another small peak around $m_{H_0}\sim 190$~GeV as new annihilation channels $A_0A_0,H^+H^-\to XX$ appear before $H_0H_0\to XX$ annihilation. Comparing the plots in \cref{fig:fig1} for the case $\lambda_{\Phi X}=10^{-15}$ with $\lambda_{\Phi X}=0.01$, we conclude that the enhancement in the relic density is due to the production of $H_0$ particles from annihilations $XX\to H_0 H_0$. However, with increasing $\lambda_{\Phi X}$, this situation changes. With increase in $\lambda_{\Phi x}$, the annihilation $\langle \sigma {\rm v}\rangle_{X X \to H_0H_0}$ also increases. However, this results in a reduction of $n_X$ number density and rise in $n_{H_0}$ number density. As a result, the overall contribution of the second term in \cref{Boltzmann1} decreases which is clearly observed in \cref{fig:fig1}. Therefore, relic density of $H_0$ (as well as enhancement in relic density) decreases with increasing $\lambda_{\Phi X}$ coupling in the region $m_{H_0}<m_{X}\sim200$~GeV and tends to decrease further in the higher mass range $m_{H_0}>m_{X}$ governed by \cref{Boltzmann3}.

A discussion is in order regarding whether inert doublet itself can obtain total DM relic abundance assisted by the vector boson. Looking into \cref{fig:fig1} one may think that it is possible as for $\lambda_{\Phi X}=0.01$, relic density of IDM is overabundant. However, it is to be noted that this is the case for $\lambda_{L}=0$. An interesting feature of IDM that distinguishes it from ordinary scalar DM is the gauge interactions which are very strong. As a result, even with $\lambda_{L}=0$ and no conversion into vector particle via direct interaction ($\lambda_{\Phi X}=0$) and Higgs portal (since $\lambda_{HX},\lambda_{L}=0$), for $m_{H_0}\geq m_{W,Z}$ we have large annihilation into gauge bosons and DM relic density is very low following the green plot shown in \cref{fig:fig1}. Now if we consider a pure IDM case with $\lambda_{L}\neq 0, \lambda_{HX}=\lambda_{\Phi X}=0$, with new annihilation channels into SM particles relic density of IDM will decrease further. However, in the Higgs portal DM case, since there is no direct interaction with the gauge sector, the annihilation of DM depends only on Higgs portal coupling and with an increase in the coupling, relic density decreases. Since in IDM, the gauge interaction dominates for $m_{H_0}>m_{W, Z}$ a small value of $\lambda_{L}=0.01$ will not affect the relic abundance very much but for a higher value of $\lambda_L=0.1$, the relic density can be even smaller due to more annihilations. Also, larger $\lambda_L$ values will eliminate most of the parameter space (for low mass IDM $m_{H_0}<500$ ~GeV) due to large direct detection cross section. However, in the presence of the vector DM candidate, it is possible to enhance the IDM relic density from conversion mechanism with new production channels $XX\rightarrow H_0 H_0$ as discussed in \cref{fig:fig1}. Now, we will discuss a case of general IDM with $\lambda_L\neq 0$ and try to investigate to what extent the relic density of IDM can be enhanced in the general situation.

\begin{figure}[!ht]
  \centering
  \includegraphics[width=\linewidth]{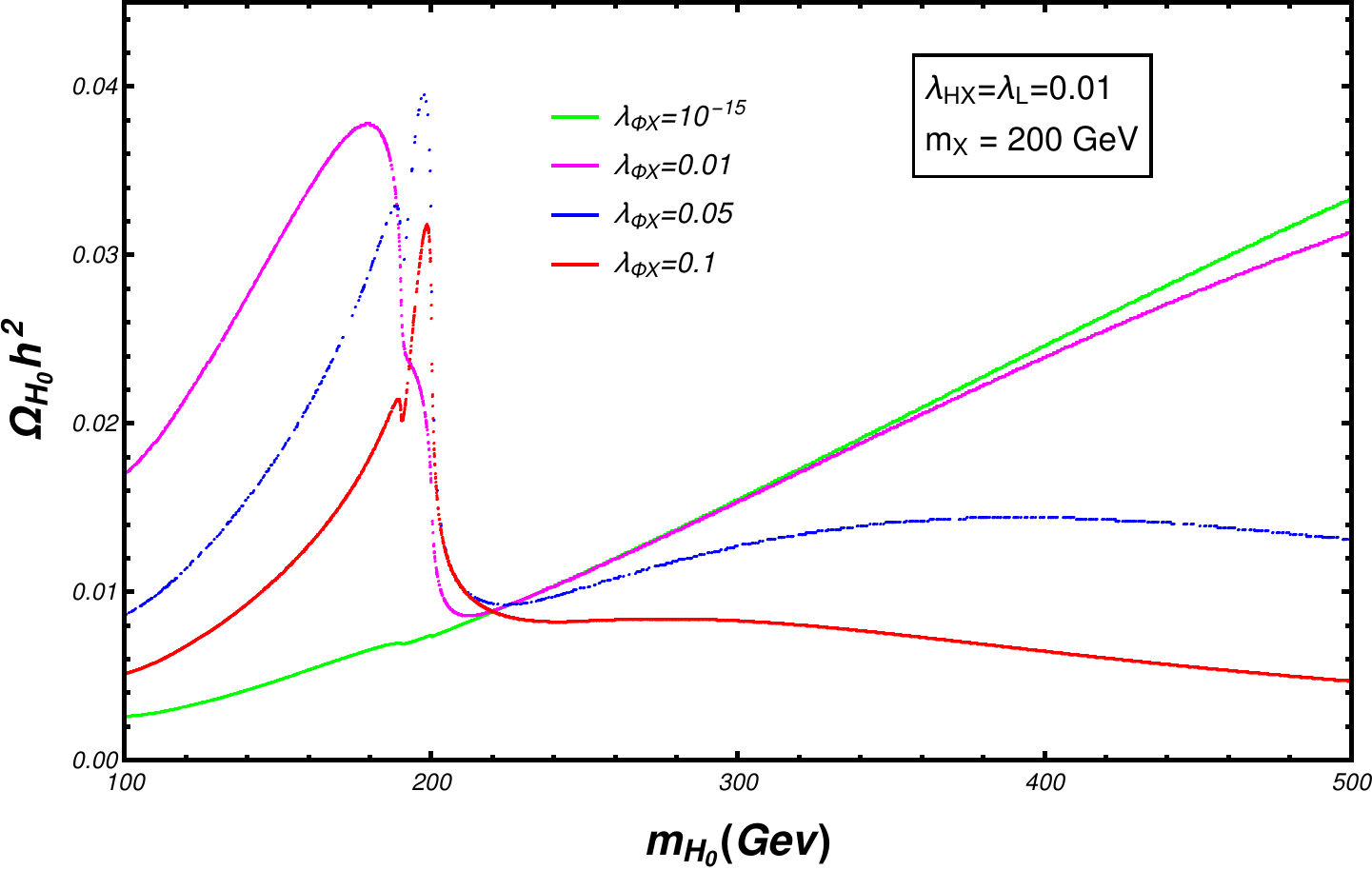}
  \caption{$\Omega_{H_0}{\rm h}^2$ vs $m_{H_0}$ for $\Delta$m$=10$~GeV. All the possible interactions (both quartic and Higgs mediated) in $H_0H_0\to SM\ SM$ and $X\ X\to H_0\ H_0$ are taken into account.}
  \label{fig:fig2}
\end{figure}

We now consider a more realistic picture than what is shown in \cref{fig:fig1}, with all possible channels of inert dark matter annihilation into consideration by taking $\lambda_L\neq0$. For this, we demonstrate our results with fixed Yukawa couplings $\lambda_L,~\lambda_{HX}=0.01$ with $\Delta m = 10$~GeV and $m_{X} = 200$~GeV for four values of $\lambda_{\Phi X}=10^{-15},0.01,0.05,0.1$.\footnote {As mentioned earlier, for larger values of $\lambda_L=0.1$, there will be large annihilations of IDM which will further decrease the IDM relic density. Moreover, larger $\lambda_L$ will also result in large direct detection cross section and reduce low mass IDM parameter space. Therefore we consider a smaller value of $\lambda_L=0.01$. A detailed discussion on $\lambda_L$ is presented later.} We show the variation in relic density of inert scalar dark matter $H_0$ with mass $m_{H_0}$ in \cref{fig:fig2}. From \cref{fig:fig2} we observe that for $\lambda_{\Phi X}=10^{-15}$, the results for inert scalar dark matter are identical to the usual inert dark matter model. However, situations change as the mixing $\lambda_{\Phi X}$ becomes large and a prominent resonance appears near the mass of $m_X$ and relic density enhances as $m_{H_0}$ approaches $m_X$. Plots with different $\lambda_{\Phi X}=0.01,0.05,0.1$ follow the similar pattern in appearing in \cref{fig:fig1}. However, due to the presence of new annihilation channels of $H_0$ (as $\lambda_L=0.01$), the enhancement in $\Omega_{H_0}{\rm h}^2$ is reduced for the mass range $m_{H_0}\leq m_X$ when compared with \cref{fig:fig1}. For the regime $m_{H_0}\leq m_X$, inert scalar particles are being produced via the annihilation of $X$ particles which compensates the annihilation of $H_0H_0$ into $W^+W^-$. Therefore although $H_0$ particles are being annihilated they are also produced at a larger rate which enhances their relic density contribution from IDM. It is to be noted that although there is a new production channel $XX \rightarrow H_0H_0$ through Higgs, its contribution is not significant and suppressed with respect to direct production depending on coupling $\lambda_{\Phi X}$. The inert scalar relic density then decreases with increasing mass $m_{H_0}>m_X$ and this effect becomes large for larger values of $\lambda_{\Phi X}$. This indicates that for the region of mass $m_{H_0}>m_X$, $H_0H_0\to XX$ annihilation becomes large and as a result relic density of IDM candidate reduces considerably. The most interesting feature is the resonance region where IDM relic density is enhanced within the 100-500~GeV regime depending on the mass of $m_X$ (which is required to be in the intermediate mass regime as well) when compared with the normal single component IDM scenario equivalent to the case $\lambda_{\Phi X}=10^{-15}$. As seen in \cref{fig:fig2}, relic density of inert dark matter candidate $H_0$ is $\Omega_{H_0}{\rm h}^2\sim 0.032$ for $m_{H_0}=200$~GeV with $\lambda_{\Phi X}=0.01$ which is much larger than $\Omega_{H_0}{\rm h}^2=7.8\times10^{-3}$ when compared with the usual single component IDM case (for $\lambda_{\Phi X}=10^{-15}$ shown in \cref{fig:fig1}). Moreover, the same order of relic density is achieved at mass $m_{H_0}\sim 500$~GeV in the normal inert doublet. We also observe that increasing $\lambda_{\Phi X}$ also reduces DM relic density $\Omega_{H_0}{\rm h}^2$ by increasing contribution of $H_0H_0\leftrightarrow XX$ channel which then starts to dominate over inert doublet annihilations in Boltzmann equation following the same pattern obtained in \cref{fig:fig1}. Therefore, one can have a significant contribution from inert DM candidate in the present scenario even within the mass range 130-500~GeV.

\begin{figure}[!ht]
  \centering
  \includegraphics[width=\linewidth]{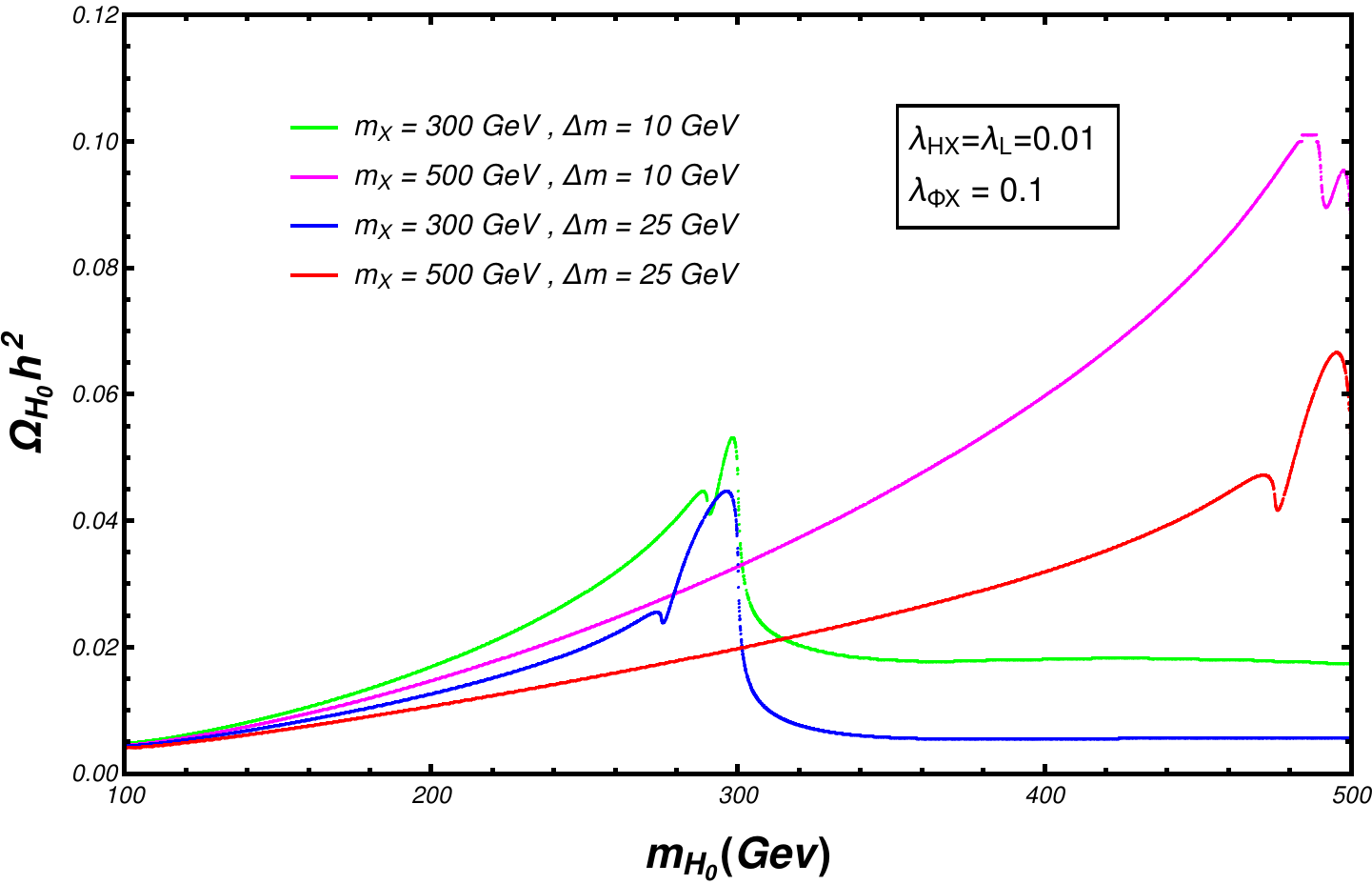}
  \caption{$\Omega_{H_0}{\rm h}^2$ vs $m_{H_0}$ for $\lambda_{\Phi X}=0.1$ as a function of $m_X$.}
  \label{fig:fig3}
\end{figure}

In another plot (\cref{fig:fig3}) we show the variations of $\Omega_{H_0}{\rm h}^2$ for the same range of $m_{H_0}$ with two different values of $m_X=300$~GeV and $500$~GeV for $\lambda_{\Phi X}=0.1$ taking $\Delta m=10$~GeV and compare the results with $\Delta m=25$~GeV for same set of parameters. The couplings $\lambda_L$ and $\lambda_{HX}$ are kept fixed at the same value 0.01. We observe that for $m_X=300$~GeV, $\Omega_{H_0}{\rm h}^2$ increases with $m_{H_0}$ until the resonance. In this regime, the annihilation $XX \to H_0H_0$ dominates which enhances $\Omega_{H_0}{\rm h}^2$ and a sharp fall appears thereafter. Similar behaviour is observed for $m_X=500$~GeV. This nature is followed for both values of $\Delta m$. However, it is to be noted that for the case $\Delta m=10$~GeV, $\Omega_{H_0}{\rm h}^2$ is larger with respect to the case when $\Delta m=25$~GeV. This is due to the fact that for smaller $\Delta m$, the contribution of co-annihilation channels in inert doublet is large which can significantly enhance relic density of inert scalar $H_0$. On the other hand for larger mass splitting, co-annihilation channels are suppressed resulting reduction of relic abundance. The spread of the peak near $m_{H_0}\sim m_{X}$ also depends on the value of mass splitting as clearly observed in \cref{fig:fig3}. The resonances peak in the plots is sharper for $\Delta m=10$~GeV and broader for $\Delta m=25$~GeV. The resonance peak near $m_{H_0}\sim m_X$ is associated with another small peak which appears as $A_0 A_0\leftrightarrow XX$ and $H^\pm H^\pm \leftrightarrow XX$ channels open up before $H_0H_0\leftrightarrow XX$. Therefore, one can conclude that depending on the mass of vector DM candidate and mass splitting $\Delta m$, $H_0$ can have a formidable contribution to total DM relic density. However from both \cref{fig:fig2,fig:fig3}, we observe that although relic abundance of inert dark matter is increased, it cannot alone satisfy the total DM relic density for dark matter and therefore the vector dark matter also contributes to the total DM relic density.

So far we have discussed how relic abundance of IDM changes with the inclusion of additional vector dark matter and found that although there is a sizeable enhancement in $H_0$ relic density, we need contribution from vector DM also. It is to be noted that vector DM can alone satisfy the DM relic density (see Ref.~\cite{Azevedo:2018oxv} and references therein). However, it is found that although portal vector DM can explain DM relic density with $\lambda_{HX}\sim 0.1$, it fails to satisfy direct detection limits for a large range of mass range up to 1 TeV (see Fig.~19 of Ref.~\cite{Azevedo:2018oxv}).  Therefore, we investigate in this work whether a low mass vector DM can be achieved to satisfy direct detection limits and also provide the required amount of relic abundance such that \cref{planck} is satisfied. If the coupling $\lambda_{HX}$ is increased further, the relic abundance of pure Higgs portal vector dark matter will fall (as $\Omega{\rm h^2} \sim 1/\langle \sigma v \rangle$), but it will be ruled out by direct detection. On the other hand for smaller $\lambda_{HX}=0.01$, direct detection can be recovered but vector DM becomes overabundant. Hence, it is interesting to study whether, in the two-component framework, relic density of vector dark matter can be reduced for small $\lambda_{HX}$.
\begin{figure}[!ht]
  \centering
  \includegraphics[width=\linewidth]{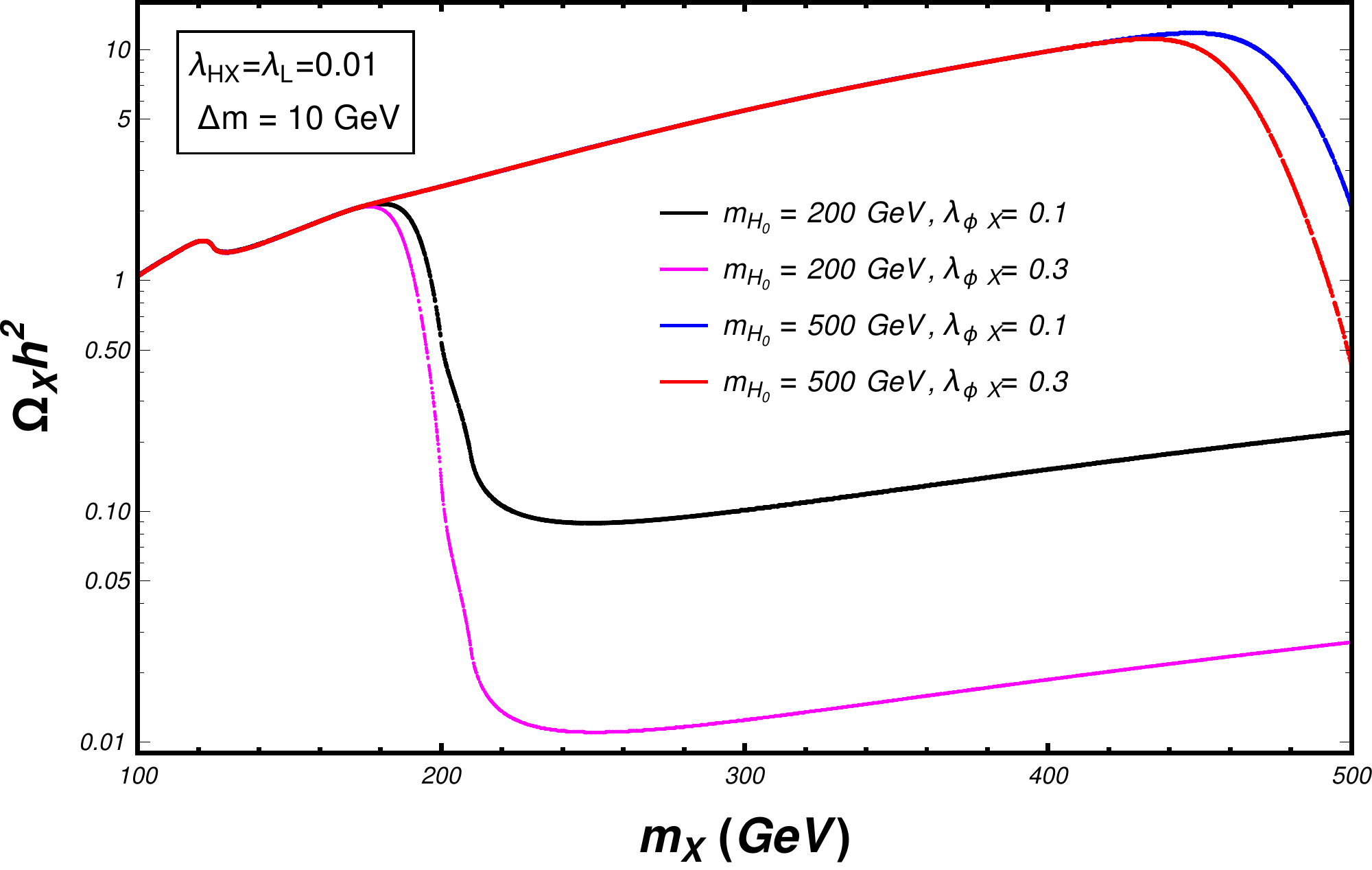}
  \caption{$\Omega_X{\rm h}^2$ vs $m_X$ for different values of \(m_{H_0}\) and $\lambda_{\Phi X}$.}
  \label{fig:fig4}
\end{figure}
In \cref{fig:fig4}, we present the variation of relic density of vector dark matter with its mass $m_X$ for two different values of $\lambda_{\Phi X}=0.1, 0.3$ and for fixed $\lambda_{L}=\lambda_{HX}=0.1$ and $\Delta m=10$~GeV. With this choice, we can directly investigate whether we have a region where vector dark matter becomes underabundant in order to satisfy the requirements of combined relic density following \cref{planck}.  From \cref{fig:fig4}, we observe that for $m_{H_0}=200$~GeV, the relic abundance of vector DM initially starts with a higher value for $m_{X}\leq m_{H_0}$. An initial drop in relic density occurs when the annihilation channel $XX\rightarrow hh$ opens for $m_X=m_h$. However, as $m_X$ increases new annihilation channel $XX\rightarrow H_0 H_0$ (when $m_X\sim m_{H_0}$) opens up  and as a result relic density of $X$ decreases. This reduction in relic density depends on the coupling $\lambda_{\Phi X}$. It can be easily observed from \cref{fig:fig4} that for $\lambda_{\Phi X}=0.3$, annihilation $XX\rightarrow H_0 H_0$ is large and relic density of vector DM falls considerably when compared with the case of $\lambda_{\Phi X}=0.1$. Similar nature of the relic density plots is observed for $m_{H_0}=500$~GeV, but in this case, the relic abundance of vector DM decreases at higher mass $m_X\sim m_{H_0}$ when the annihilation becomes kinematically allowed. Therefore, the low mass region of vector dark matter which was ruled out by direct detection in standard Higgs portal scenario becomes accessible in the two-component framework and also opens the window for the study of multicomponent dark matter.

\begin{figure}[!ht]
  \centering
  \includegraphics[width=\linewidth]{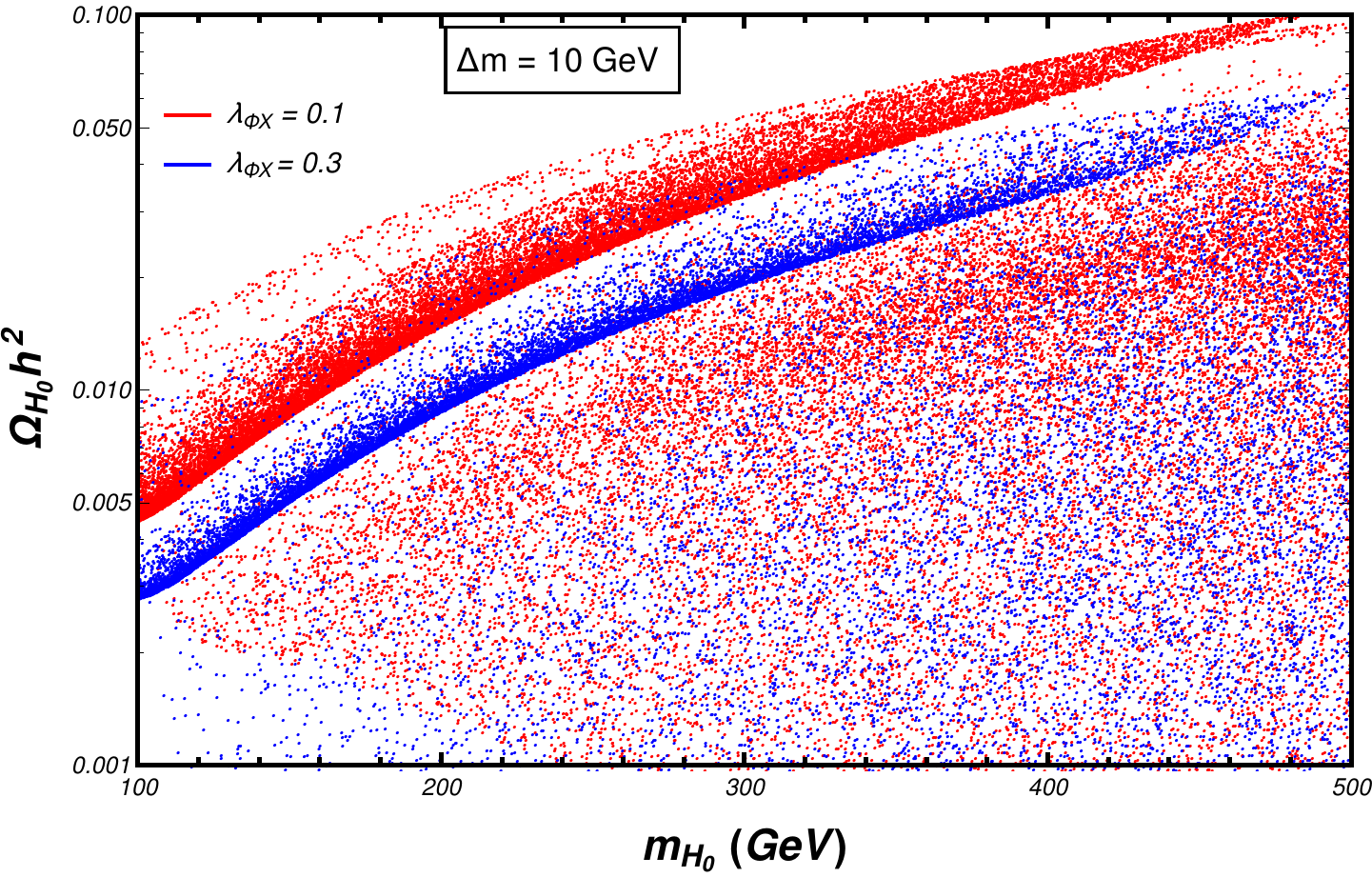}
  \caption{$\Omega_{H_0}{\rm h}^2$ vs $m_{H_0}$ as a function of $\lambda_{\Phi X}$ for $\Delta m=10$~GeV.}
  \label{fig:fig5}
\end{figure}

Now we present our results for the overall range of parameter space described at the beginning of this Section. In \cref{fig:fig5} we depict a scatter plot of $m_{H_0}$ against $\Omega_{H_0}{\rm h}^2$ for two values of $\lambda_{\Phi X}=0.1,0.3$ varying both $m_{H_0}$ and $m_{X}$ from 100-500~GeV with $\Delta m=10$~GeV. We have considered the same values of $\lambda_L,\lambda_{HX}=0.01$ as stated earlier. We observe a large spread in relic density $\Omega_{H_0}{\rm h}^2$ with $m_{H_0}$. We also observe in \cref{fig:fig5} that increasing $\lambda_{\Phi X}$ results in decrease in the relic density $\Omega_{H_0}{\rm h}^2$ which directly follows the nature of plots in \cref{fig:fig1} and \cref{fig:fig2}. For the chosen $\Delta m=10$~GeV, we observe that the relic density $\Omega_{H_0}{\rm h}^2$ attains a maximum $~0.1$ for $m_{H_0}\sim500$~GeV for $\lambda_{\Phi X}=0.1$ and for $\lambda_{\Phi X}=0.3$ it can reach values up to $0.05$. This is due to increase in $\lambda_{\Phi X}$ which increases the $2\leftrightarrow 2$ annihilations between dark sector particles as stated earlier.

\begin{figure}[!ht]
  \centering
  \includegraphics[width=\linewidth]{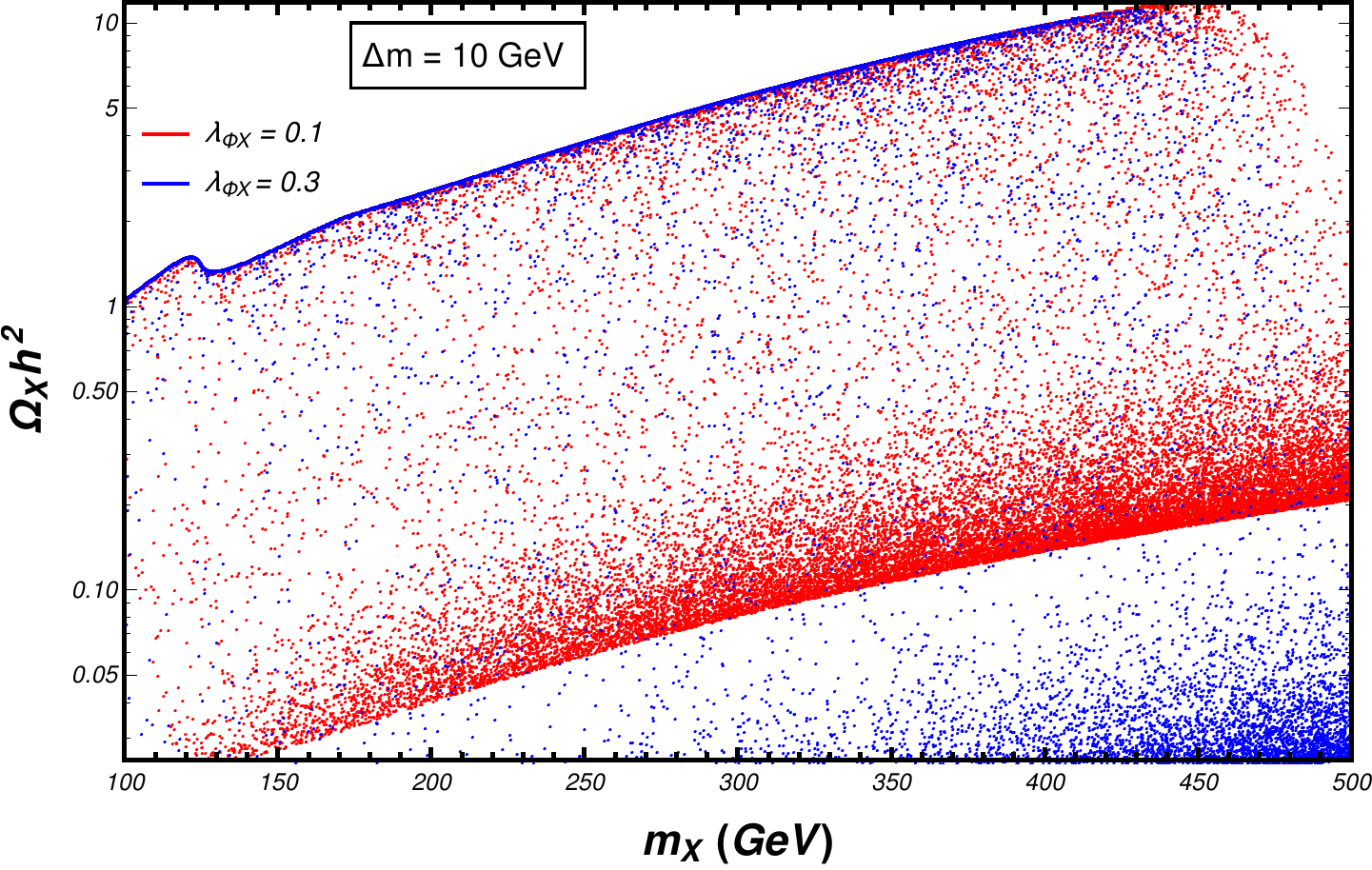}
  \caption{$\Omega_{X}{\rm h}^2$ vs $m_{X}$ as a function of $\lambda_{\Phi X}$ for $\Delta m=10$~GeV.}
  \label{fig:fig6}
\end{figure}

\begin{figure}[!ht]
  \centering
  \includegraphics[width=\linewidth]{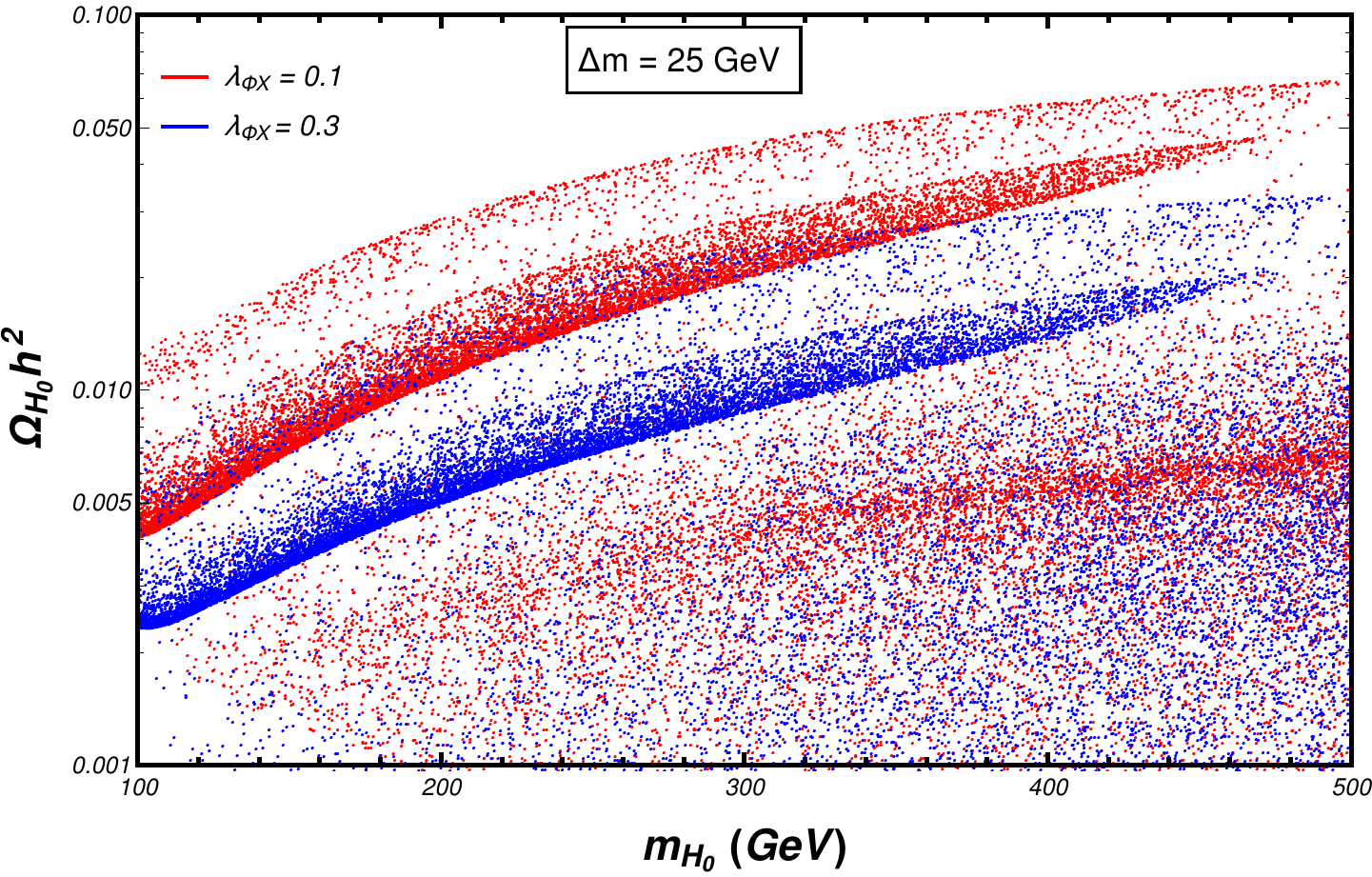}
  \caption{$\Omega_{H_0}{\rm h}^2$ vs $m_{H_0}$ as a function of $\lambda_{\Phi X}$ for $\Delta m\!=\!25$~GeV.}
  \label{fig:fig7}
\end{figure}

\begin{figure}[!ht]
  \centering
  \includegraphics[width=\linewidth]{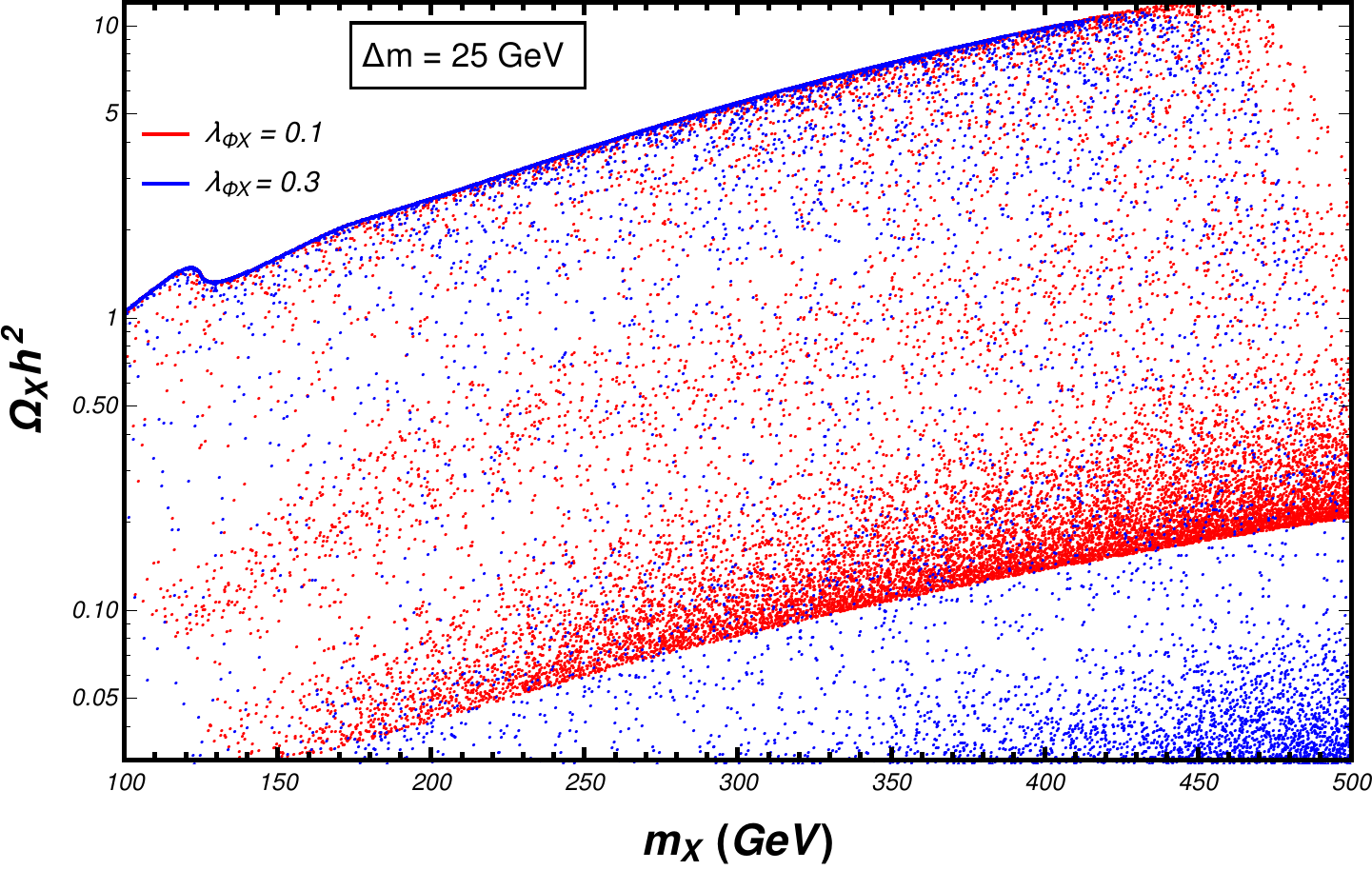}
  \caption{$\Omega_{X}{\rm h}^2$ vs $m_{X}$ as a function of $\lambda_{\Phi X}$ for $\Delta m=25$~GeV.}
  \label{fig:fig8}
\end{figure}

In \cref{fig:fig6} we show the variation of $m_X$ with $\Omega_{X}{\rm h}^2$ for the same range of parameters. Here we also observe a similar increase in relic density with $m_X$ as obtained for scalar DM candidate. However, a large range of parameter space is ruled out for as $\Omega_{X}{\rm h}^2$ becomes overabundant. A small drop in $\Omega_X{\rm h^2}$ near $m_X\sim m_h$ represents the new annihilation channels $XX\to hh$. Comparing \cref{fig:fig4} with \cref{fig:fig6}, it can be easily observed that region of upper and lower part of the envelope in \cref{fig:fig6} follows directly from \cref{fig:fig4} as the mass $m_{H_0}$ is varied from 100 to 500~GeV. For smaller values of $m_X$, as shown in \cref{fig:fig4} a drop in $\Omega_X{\rm h^2}$ occurs which creates the lower envelope while for higher values of $m_X$ the reduction in relic density appears later creating the upper envelope of the scatter plot. Similar plots for variations for $\Omega_{H_0}{\rm h}^2$ ($\Omega_{X}{\rm h}^2$) against $m_{H_0}$ ($m_X$) with $\Delta m=25$~GeV are shown in \cref{fig:fig7,fig:fig8}. It is interesting to observe that for increased mass splitting between $H_0$ and $A_0$ and for $\lambda_{\Phi X}=0.1$, the maximum relic density obtained for the inert DM (${\Omega_{H_0}{\rm h}^2}\sim0.05$) is almost half when compared with \cref{fig:fig5}. This is due to the fact that as we increase the mass splitting, the contributions from co-annihilation channels will reduce. The plot in \cref{fig:fig7} depicts same nature of \cref{fig:fig5} as we change $\lambda_{\Phi X}$. However, for the vector DM candidate $X$, there is no such effect and the corresponding \cref{fig:fig8} remains almost similar to \cref{fig:fig6} with no significant change. Both \cref{fig:fig6} and \cref{fig:fig8} exactly follows the nature of \cref{fig:fig4} discussed earlier. Moreover, the choice of $\Delta m$ does not affect the relic density of $X$ as it is related to inert dark matter. However, we will show that although it doesn't affect the relic density $\Omega_{X}{\rm h}^2$, mass splitting $\Delta m$ can have a significant effect when the total relic density of the multicomponent dark sector is taken into account.

We would like to mention here that a simple scenario for scalar and vector DM without mixing \(\lambda_{\Phi X}\) is completely different from inert doublet and vector DM. First of all, inert doublet has direct gauge interaction but singlet scalar DM can interact with gauge boson through Higgs portal only. Also in the case of singlet scalar DM (S), there is only one conversion channel \(SS \leftrightarrow XX\) (via quartic coupling \(\lambda_{\Phi X}\) and Higgs mediation). But for inert doublet, there are many other annihilation channels, such as $H_0H_0, A_0A_0, H^+H^-\leftrightarrow XX$ (the analytical expressions for the annihilation cross section are given in the \cref{app:appA}), that can affect the result. Furthermore, it also depends on the mass splitting between $H_0,A_0$ etc. Therefore, results with inert doublet and vector DM are substantially different from the study of singlet scalar + vector DM discussed in Ref.~\cite{Bian:2013wna}. This can be easily understood from \cref{fig:fig3}. In the present multicomponent DM, depending on mass splitting we have two distinct peaks in \cref{fig:fig3}, small peaks due to $A_0A_0,H^+H^- \leftrightarrow XX$ and a large peak due to $H_0H_0 \leftrightarrow XX$. In the case of singlet + vector DM, there will be only one such peak only for $ SS \leftrightarrow XX$ as there is no other annihilation possible.

\begin{figure}[!ht]
  \centering
  \includegraphics[width=\linewidth]{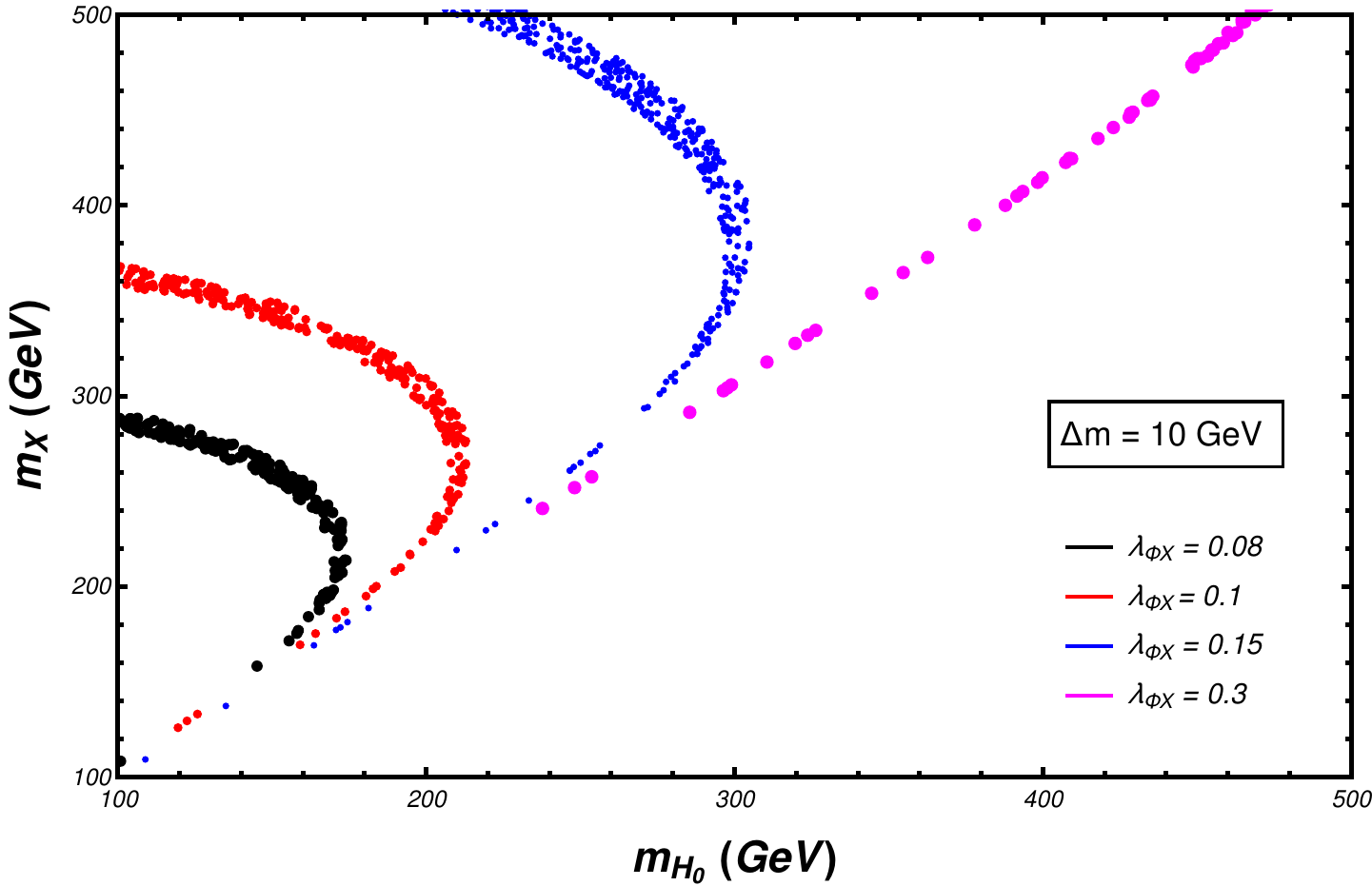}
  \caption{$m_X$ vs $m_{H_0}$ for $\Delta m=10$~GeV as a function of $\lambda_{\Phi X}$.}
  \label{fig:fig9}
\end{figure}
\begin{figure}[!ht]
  \centering
  \includegraphics[width=\linewidth]{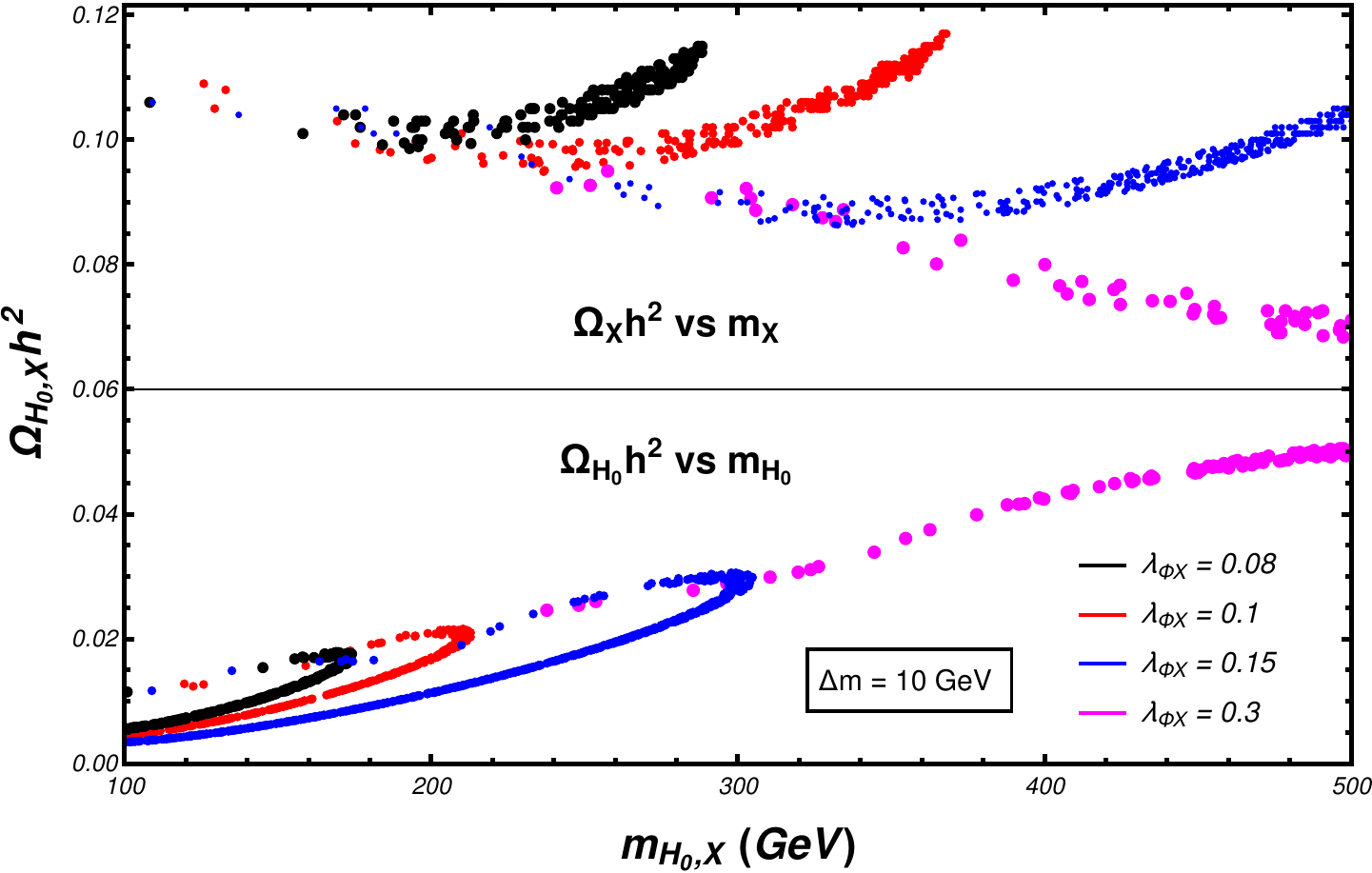}
    \caption{$\Omega_{X}{\rm h}^2$ vs $m_{X}$,$\Omega_{H_0}{\rm h}^2$ vs $m_{H_0}$ for $\Delta m=10$~GeV.}
  \label{fig:fig10}
\end{figure}

In \cref{fig:fig9}, we present the allowed ranges of dark matter masses in $m_{H_0}-m_X$ plane for four values of $\lambda_{\Phi X} = 0.08, 0.1, 0.15, 0.3$ considering $\lambda_L=\lambda_{HX}=0.01$ with $\Delta m=10$~GeV. We vary the masses of dark matter candidates in the range $100~{\rm~GeV}\leq m_{H_0,X}\leq500~\rm{GeV}$ as mentioned in the beginning of this section and use the condition expressed in \cref{planck} for total DM relic abundance. From the plots in \cref{fig:fig9} we observe that apart from the region when $m_{H_0}\sim m_X$, there exists another regime with $m_{X}>m_{H_0}$. However this nature disappears as we increase the coupling $\lambda_{\Phi X}$. From the \cref{fig:fig9}, we notice that in case of $\lambda_{\Phi X}=0.1$, the maximum mass $m_{H_0}\sim220$~GeV and then it tends to decrease with increasing $m_X$ while $m_{H_0}$ reaches a maximum value about 300~GeV for $\lambda_{\Phi X}=0.15$.

In order to explain the nature of these plots in \cref{fig:fig10} we present the variation of $\Omega_{H_0,X}{\rm h^2}$ vs $m_{H_0,X}$ for the same set of parameters considered in \cref{fig:fig9}. From \cref{fig:fig10}, it can be observed that for $\lambda_{\Phi X}=0.1$, initially the relic density $\Omega_{X}{\rm h}^2$ tends to decrease with increasing $m_X$. In this regime, the contribution of inert dark matter $H_0$ as well as the mass of $H_0$ also increases. This can easily be checked by adding up the relic density $\Omega_{H_0}{\rm h}^2$ with $\Omega_{X}{\rm h}^2$ that provide the required total DM relic abundance. This corresponds to the regime $m_X\sim m_{H_0}$, the lower half of the semicircular arc in \cref{fig:fig9} for $\lambda_{\Phi X}=0.1$ that continues till $m_X\sim m_{H_0}\sim220$~GeV. For larger values of $m_X>220$~GeV, the relic density $\Omega_{X}{\rm h}^2$ tends to increase while the contribution of $H_0$ in total DM relic decreases which also results in a reduction of the mass of $H_0$. This regime corresponds to the upper part of the semicircular allowed region shown in \cref{fig:fig9}. This can also be justified by looking into \cref{fig:fig6} where $\Omega_{X}{\rm h}^2$ becomes overabundant near $m_X\sim 380$~GeV for $\lambda_{\Phi X} = 0.1$. This indicates that with the increase in $m_X$, the contribution of the other candidate must be small as $\Omega_{X}{\rm h}^2$ approaches $\Omega_{DM}{\rm h^2}$. Similar conclusion can be drawn for $\lambda_{\Phi X}=0.08,0.15$ from \cref{fig:fig9,fig:fig10}. However, with increasing values of $\lambda_{\Phi X}=0.3$ such feature disappears as observed in \cref{fig:fig9}. We have found that with the present two component scenario, the relic density contribution of $H_0$ can be enhanced significantly. For example, in \cref{fig:fig10}, the relic density of $H_0$ having mass $m_{H_0}=400$~GeV when calculated for $\lambda_{\Phi X}=0.3$ is about $\Omega_{H_0}{\rm h}^2=0.042$ with respect to the usual value $\Omega_{H_0}{\rm h}^2=0.0248$ without influence of other DM candidate.

\begin{figure}[!ht]
  \centering
  \includegraphics[width=\linewidth]{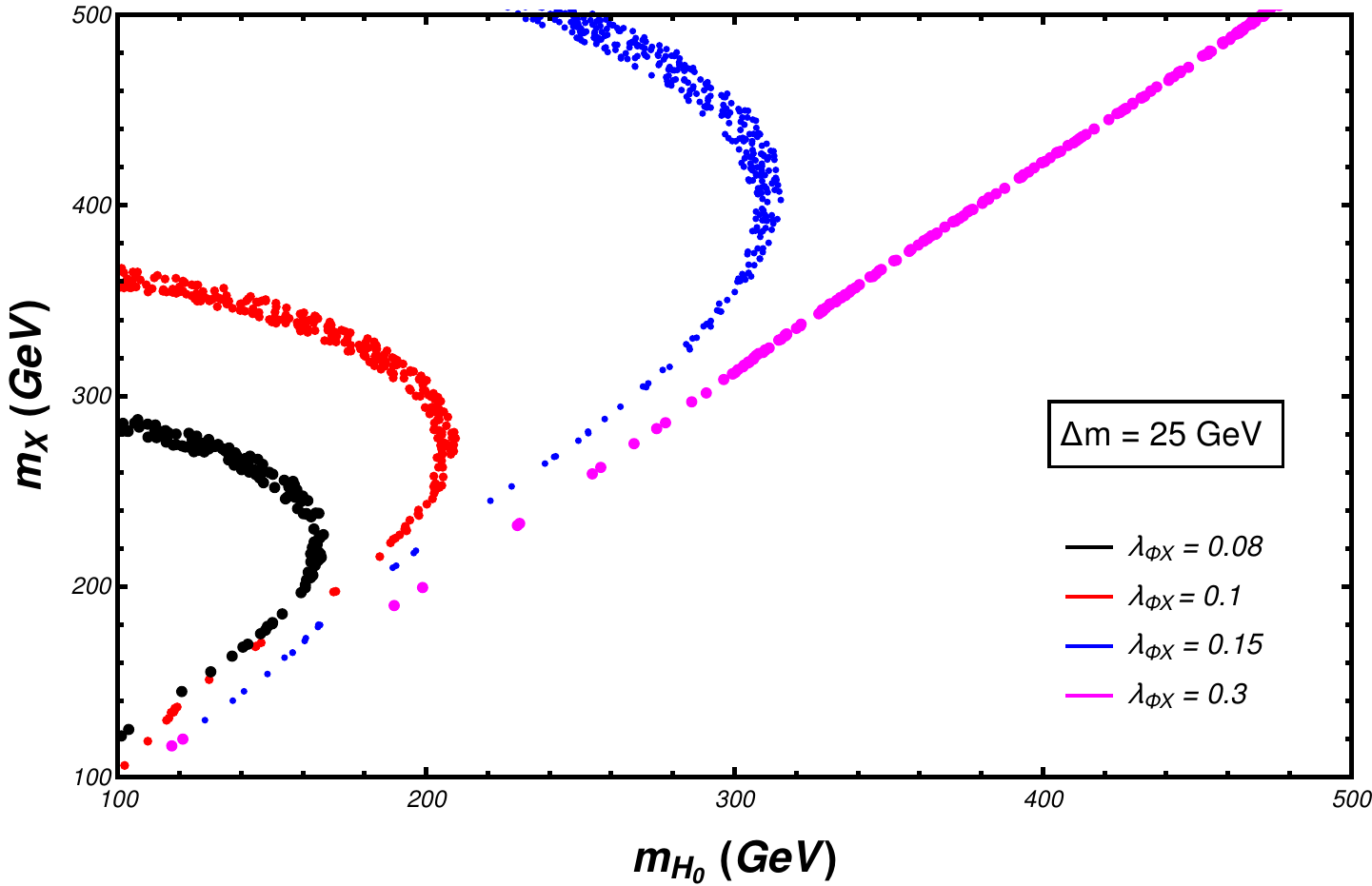}
  \caption{$m_X$ vs $m_{H_0}$ for $\Delta m=25$~GeV as a function of $\lambda_{\Phi X}$.}
  \label{fig:fig11}
\end{figure}
\begin{figure}[!ht]
  \centering
  \includegraphics[width=\linewidth]{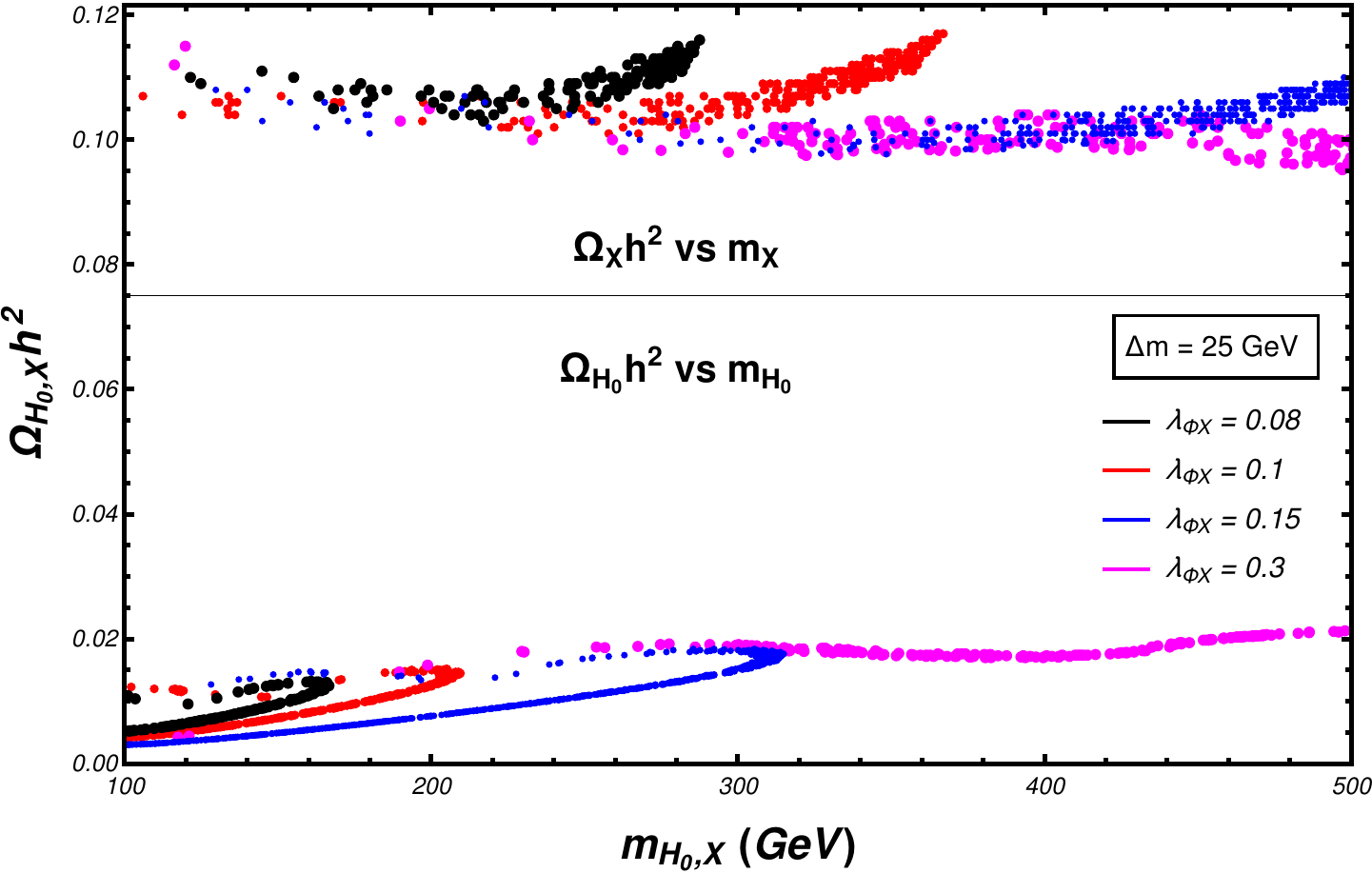}
  \caption{$\Omega_{X}{\rm h}^2$ vs $m_{X}$,$\Omega_{H_0}{\rm h}^2$ vs $m_{H_0}$ for $\Delta m=25$~GeV.}
  \label{fig:fig12}
\end{figure}

Similar plots are shown in \cref{fig:fig11,fig:fig12} with mass splitting $\Delta m=25$~GeV. \cref{fig:fig9,fig:fig11} depict similar nature and as we increase the value $\lambda_{\Phi X}$, the relation between $m_X$ and $m_{H_0}$ becomes more linear $m_X\sim m_{H_0}$. Although \cref{fig:fig11} is almost same as \cref{fig:fig9} when compared for different values of $\lambda_{\Phi X}$, due to the larger mass splitting contribution of $\Omega_{H_0}{\rm h}^2$ to total dark matter relic density is smaller when compared with the case $\Delta m=10$~GeV. This is shown in \cref{fig:fig12} and is obvious as for larger $\Delta m$ values contribution in $\Omega_{H_0}{\rm h}^2$ from co-annihilation channels are suppressed. However, still, there is sufficient contribution from the inert dark matter in total DM relic density. It can be observed that the value of $\Omega_{H_0}{\rm h}^2=0.021$, for $m_{H_0}=500$~GeV with $\Delta m=25$~GeV when $\lambda_{\Phi X}=0.3$. In the absence of the multicomponent scenario, the contribution of inert doublet candidate becomes $\Omega_{H_0}{\rm h}^2=0.007$. Therefore, the extension of the inert doublet with a vector dark matter can provide a successful multicomponent dark matter scenario where the contribution of the inert candidate can be enhanced considerably. However such properties are prominent for larger $\lambda_{\Phi X}$ values and when $m_{X}\sim m_{H_0}$. For smaller $\lambda_{\Phi X}$, the contribution of inert doublet increases up to a certain value satisfying the relation $m_X\sim m_{H_0}$ and then decrease with an increase in the mass of $m_X$. In such case, the inert doublet may have very small relic density (even when $m_X\geq m_{H_0}$) and most of the DM relic abundance is obtained from the other candidate $X$. Overall, from \cref{fig:fig9} and \cref{fig:fig11} we can conclude that $m_{H_0}\leq m_{X}$ region in $m_{H_0}$ vs $m_{X}$ plane is allowed in the present multicomponent model depending on the choice of $\lambda_{\Phi X}$.

\begin{description}[leftmargin=0pt,labelindent=0pt]
\item[{\bf Direct and Indirect detection of DM candidates:}]
As mentioned earlier, from the study of vector dark matter~\cite{Arcadi:2019lka}, it is observed that although a vector dark matter can satisfy DM relic density with $\lambda_{HX}\sim 0.1$ for $m_X\sim100-500$~GeV, it is ruled out by direct detection limits from XENON1T. For a smaller value of $\lambda_{HX}=0.01$, it can be easily found that vector dark matter is in agreement with XENON1T limit for $m_X\geq 140$~GeV. However, for portal vector dark matter, relic density will be large as annihilation cross section becomes small. On the other hand in the two-component scenario, as shown in \cref{fig:fig4} when $m_X>m_{H_0}$, $\Omega_{X}{\rm h^2}$ decreases due to new annihilation channels into an inert doublet. From \cref{fig:fig10,fig:fig11,fig:fig12}, we also observe that contribution to the total DM relic density from vector dark matter in the multicomponent framework is larger compared to the inert doublet. Therefore, conservatively assuming $\Omega_{X}{\rm h^2}\sim \Omega_{DM}{\rm h^2}$, $\lambda_{HX}=0.01$ allows us to study the low mass vector dark matter with $m_X>140$~GeV which was excluded earlier.
\end{description}

Similar to the case of vector dark matter, the choice of $\lambda_L=0.01$ for inert doublet dark matter is not arbitrary. Different collider searches for inert doublet dark matter also consider the limit on $\lambda_{L}$ (also known as $\lambda_{345}$). Collider study of inert doublet including the direct detection limits from LUX excludes larger values of $\lambda_L$ for $m_{H_0}\sim 100 $~GeV and for $m_{H_0}=500$~GeV, maximum allowed value of $\lambda_L=0.4$~\cite{Ilnicka:2015jba,Ilnicka:2018def}. Using the direct detection bound from XENON1T, the allowed limit on $\lambda_L$ is further reduced. From~\cite{Ilnicka:2018def} (see Fig.~7 in reference mentioned), it is found that for $m_{H_0}\sim 100$~GeV $\lambda_L\leq 0.01$ and for $m_{H_0}=500$~GeV, $\lambda_L\leq 0.1$ is allowed. Therefore, in order to study inert doublet in the mass range $m_{H_0}\sim100-500$~GeV, we consider a value of $\lambda_L=0.01$.

As mentioned in the beginning of this \cref{sec:results}, we considered $\lambda_{L}=\lambda_{H X}=0.01$ and the direct detection cross section for both the candidates can be obtained from \cref{scalardd,vectordd}. Therefore, the direct detection cross section is further reduced by the scaling $r_i,~i=H_0,X$. We have found that, with the present choice of $\lambda_{L}=0.01$, the regions plotted for $H_0$ in \cref{fig:fig1,fig:fig2,fig:fig3,fig:fig5,fig:fig7,fig:fig9,fig:fig10,fig:fig11,fig:fig12}\footnote{There will not be any direct detection for \cref{fig:fig1} as $\lambda_{L}\!=\!\lambda_{H X}=0$.} with $100~{\rm~GeV}\leq m_{H_0}\leq 500$~GeV remains within the direct detection cross section limits obtained from XENON1T~\cite{Aprile:2018dbl}. Similarly for $\lambda_{H X}=0.01$ in case of the vector boson dark matter, we find that all the regions within the mass range $140~{\rm~GeV}\leq m_{X}\leq 500$~GeV plotted in \cref{fig:fig6,fig:fig8,fig:fig9,fig:fig10,fig:fig11,fig:fig12} is in agreement with the XENON1T bound. This limit is obtained by simply assuming $r_X\sim 1$ in \cref{vectordd} since in both \cref{fig:fig10,fig:fig12} we observe that for masses $m_{H_0,X}\leq 200$~GeV, relic density of IDM is small with respect to the vector dark matter. Therefore, all the allowed points with $m_{X}<140$~GeV satisfying \cref{planck} in \cref{fig:fig9,fig:fig10,fig:fig11,fig:fig12} are excluded by XENON1T bound on vector dark matter. This leaves an allowed parameter space in $m_{H_0}-m_{X}$ plane (shown in \cref{fig:fig9,fig:fig11}) satisfying the conditions i) $100~{\rm~GeV}\leq m_{H_0}\leq 500$~GeV, ii) $140~{\rm~GeV}\leq m_{X}\leq 500$~GeV and iii)$m_{H_0}\leq m_{X}$ for fixed $\lambda_L,\lambda_{HX}, \Delta m$ considering $\lambda_{\Phi X}$ as the only variable. We use these limits  for further studies involving collider searches for dark matter candidates in the next section. Of course one can avoid direct detection bound on vector dark matter by assuming a smaller value of $\lambda_{HX}$, thus consistent with XENON1T bound. But that would not affect the phenomenology of the model leaving $m_{H_0}\leq m_{X}$ as the only condition to restrain the model parameter space.

\begin{table}[!ht]
\begin{tabular}{|c |c |c |c |c |c |c |c |}
\hline
 \multicolumn{2}{|c|}{}
             & \multicolumn{2}{c|}{Indirect detection}                            & \multicolumn{2}{c|}{Direct detection} \\ \hline
 \(m_X\)
     & \multirow{2}{*}{\(\lambda_{HX}\)}
             & \(\braket{\sigma v}_{\rm X}\)
                                                & \(\braket{\sigma v}_\text{\tiny Fermi-LAT}\)
                                                                                  & \multirow{2}{*}{\(\sigma_{X} [\rm cm^2]\)}
                                                                                    &  \(\sigma_{\text{\tiny XENON1T}}\)\\
\tiny{[GeV]}
     &       & \([\rm cm^3/s]\)                 & \([\rm cm^3/s]\)                & &  \([\rm cm^2]\)\\ \hline\hline
 100 & 0.001 & \(2.28\!\!\times\!\! 10^{-29}\)  & \multirow{4}{*}{\(3.0\!\!\times\!\! 10^{-26}\)}
                                                  & \(3.46\!\!\times\!\! 10^{-48}\) & \multirow{4}{*}{\(1.5\!\!\times\!\! 10^{-46}\)}\\ \cline{1-3}\cline{5-5}
 100 & 0.01  & \(2.27 \!\!\times\!\! 10^{-27}\) & & \(3.46\!\!\times\!\! 10^{-46}\) & \\ \cline{1-3}\cline{5-5}
 100 & 0.02  & \(9.0 \!\!\times\!\! 10^{-27}\)  & & \(1.38\!\!\times\!\! 10^{-45}\) & \\ \cline{1-3}\cline{5-5}
 100 & 0.03  & \(2.0 \!\!\times\!\! 10^{-26}\)  & & \(3.11\!\!\times\!\! 10^{-45}\) & \\ \hline
 500 & 0.001 & \(1.25\!\!\times\!\! 10^{-28}\)  & \multirow{4}{*}{\(1.0\!\!\times\!\! 10^{-25}\)}
                                                  & \(1.48\!\!\times\!\! 10^{-49}\) & \multirow{4}{*}{\(4.0\!\!\times\!\! 10^{-46}\)}\\ \cline{1-3}\cline{5-5}
 500 & 0.01  & \(1.37\!\!\times\!\! 10^{-26}\)  & & \(1.3 \!\!\times\!\! 10^{-47}\) & \\ \cline{1-3}\cline{5-5}
 500 & 0.1   & \(9.1\!\!\times\!\! 10^{-30}\)   & & \(1.27 \!\!\times\!\! 10^{-45}\)& \\ \cline{1-3}\cline{5-5}
 500 & 1.0   & \(5.3 \!\!\times\!\! 10^{-26}\)  & & \(3.8 \!\!\times\!\! 10^{-44}\) & \\ \hline
\end{tabular}
  \caption{Bound on \(\lambda_{HX}\) from direct and indirect detection. $m_{H_0}=100$~GeV, $\lambda_{\Phi X}$=0.1, $\lambda_{L}$=0.01.}
  \label{tab:tab1a}
\end{table}
\begin{table}[!ht]
\begin{tabular}{|c |c |c |c |c |c |c |c |}
\hline
 \multicolumn{2}{|c|}{}
             & \multicolumn{2}{c|}{Indirect detection}                            & \multicolumn{2}{c|}{Direct detection} \\ \hline
 \(m_{H_0}\)
     & \multirow{2}{*}{\(\lambda_L\)}
             & \(\braket{\sigma v}_{\rm H_0}\)
                                                & \(\braket{\sigma v}_\text{\tiny Fermi-LAT}\)
                                                                                  & \multirow{2}{*}{\(\sigma_{H_0} [\rm cm^2]\)}
                                                                                    &  \(\sigma_{\text{\tiny XENON1T}}\)\\
\tiny{[GeV]}
     &       & \([\rm cm^3/s]\)                 & \([\rm cm^3/s]\)                & &  \([\rm cm^2]\)\\ \hline\hline
 100 & 0.001 & \(4.47\!\!\times\!\! 10^{-28}\)  & \multirow{4}{*}{\(3.0\!\!\times\!\! 10^{-26}\)}
                                                                                  & \(7.2\!\!\times\!\! 10^{-50}\)
                                                                                    & \multirow{4}{*}{\(1.5\!\!\times\!\! 10^{-46}\)}\\ \cline{1-3}\cline{5-5}
 100 & 0.01  & \(4.14 \!\!\times\!\! 10^{-28}\) & & \(7.2\!\!\times\!\! 10^{-48}\)
                                                                                    & \\ \cline{1-3}\cline{5-5}
 100 & 0.1   & \(2.18 \!\!\times\!\! 10^{-28}\) & & \(4.13\!\!\times\!\! 10^{-46}\) & \\ \cline{1-3}\cline{5-5}
 100 & 1.0   & \(2.24 \!\!\times\!\! 10^{-29}\) & & \(8.48\!\!\times\!\! 10^{-45}\) & \\ \hline
 500 & 0.001 & \(1.81\!\!\times\!\! 10^{-33}\)  & \multirow{4}{*}{\(1.0\!\!\times\!\! 10^{-25}\)}
                                                  & \(1.25\!\!\times\!\! 10^{-50}\) & \multirow{4}{*}{\(4.0\!\!\times\!\! 10^{-46}\)}\\ \cline{1-3}\cline{5-5}
 500 & 0.01  & \(1.7\!\!\times\!\! 10^{-29}\)   & & \(1.23\!\!\times\!\! 10^{-48}\) & \\ \cline{1-3}\cline{5-5}
 500 & 0.1   & \(9.1\!\!\times\!\! 10^{-30}\)   & & \(1.0 \!\!\times\!\! 10^{-46}\)& \\ \cline{1-3}\cline{5-5}
 500 & 1.0   & \(4.26 \!\!\times\!\! 10^{-31}\) & & \(3.83\!\!\times\!\! 10^{-45}\)  & \\ \hline
\end{tabular}
  \caption{Bound on \(\lambda_L\) from direct and indirect detection. $m_X=600$~GeV, $\lambda_{\Phi X}$=0.1, $\lambda_{HX}$=0.01.}
  \label{tab:tab1b}
\end{table}

The bounds on the upper and lower limit of couplings $\lambda_L$ and $\lambda_{HX}$, obtained from direct and indirect detection experiments, are given in Table \ref{tab:tab1a} and  \ref{tab:tab1b} respectively. It is self-explanatory that our choice of parameters are consistent with these constraints, with the comment that $m_X \le 140$~GeV is excluded by direct detection limits, as already mentioned earlier which can also be verified from these Tables. While direct detection cross sections are purely Higgs mediated for both the DM candidates and therefore strongly depend on the Higgs portal couplings $\lambda_{L}$ and $\lambda_{HX}$(see \cref{scalardd,vectordd} for expression), indirect detection cross sections may not have similar dependence on the couplings, because for $H_0$, the dominant channel is annihilation into gauge pairs, which cannot be controlled by $\lambda_L$. However, indirect detection channels for X will depend on $\lambda_{HX}$ since it is the only portal through which it can interact with SM. Note that $\sigma_{H_0}~[\sigma_X]$ mentioned in the above tables are actually $\sigma_{{H_0}\rm\ SM \rightarrow {H_0}\rm\ SM}\times (\Omega_{H_0}/\Omega_{\rm DM})~[\sigma_{{X}\rm\ SM \rightarrow {X}\rm\ SM}\times (\Omega_{X}/\Omega_{\rm DM})]$, similarly, \(\braket{\sigma v}_{\rm {H_0}}\) and \(\braket{\sigma v}_{\rm X}\) are obtained by multiplying the total indirect annihilation cross sections for individual DM~DM~$\rightarrow$~SM~SM  multiplied by the fractions $(\Omega_{H_0}/\Omega_{\rm DM})^2$ and $(\Omega_{X}/\Omega_{\rm DM})^2$ respectively. The direct detection experimental bound is obtained from XENON1T~\cite{Aprile:2018dbl} and indirect bound is taken from Fermi-LAT data~\cite{Fermi-LAT:2016uux}. It is to be noted that in the mass range $100 {\rm~GeV}\leq m_{X,H_0}\leq 500$~GeV both the dark matter candidates annihilate into gauge bosons dominantly. The Fermi-LAT limit on the annihilation cross section is on DM annihilating directly into $b\bar{b}$. Both cross sections \(\braket{\sigma v}_{\rm {H_0}}\) and \(\braket{\sigma v}_{\rm X}\) are found to be much smaller with respect to Fermi-LAT limit and therefore consistent with the bounds from indirect detection since secondary annihilation cross section into $b{\bar b}$ pair will be even smaller. Upper limits  on $H_0H_0\rightarrow W^+W^-$ cross section is calculated in Ref.~\cite{Garcia-Cely:2015ysa} considering emission of gamma ray in dwarf galaxies and also anti-proton flux. We found that the annihilation cross section of  $H_0H_0\rightarrow W^+W^-$ is smaller in the present multicomponent dark matter model compared to the observed limits in Ref.~\cite{Garcia-Cely:2015ysa}.

\section{Dark Matter signatures at the LHC}
\label{sec:collider}
In this section, we will discuss the issues pertaining to the collider signatures of the dark matter specific to our model. We specifically focus on the \emph{dilepton plus missing transverse energy} channel ($2\ell + \slashed{E}_T$, $\ell = e, \mu$) in the present study. We choose at least 2 leading $p_T$ leptons in the final states irrespective of their charge. Among the two DM candidates in our model, the vector DM, $X$, is produced only in the association of the SM Higgs boson, $h$, because it belongs to a separate $Z_2'$ symmetry. As we are aware that the branching ratio of $h \to 2\ell$ is negligible, $X$ does not give any significant signal cross section in the present LHC environment. Hence the collider signature of the DM here is entirely dictated by $H_0$. Previously, Refs.~\cite{Dolle:2009ft,Belanger:2015kga} studied the DM signature at the LHC for $2\ell + \slashed{E}_T$ channel for IDM alone whereas Refs.~\cite{Bhattacharya:2018cgx} studied the same signature in the context of multiparticle DM model consisting of IDM plus scalar. In a separate study, the authors of Ref.~\cite{Datta:2016nfz} studied the multilepton channel with more than 2 leptons in the final state. Here we will not address such channels as the signal cross section will decrease significantly for them.

For this study, we first generate our model file which can be used in a Monte Carlo event generator. We build our model file using {\texttt FeynRules}~\cite{Alloul:2013bka}. This model file is then used for generation of events through {\texttt MadGraph5}~\cite{Alwall:2014hca}. We further use {\texttt Pythia 8}~\cite{Sjostrand:2014zea} for showering, fragmentation and hadronization to get {.\scshape hepmc} files. These {.\scshape hepmc} files are then processed for fast simulation in detector studies through {\texttt Delphes}~\cite{deFavereau:2013fsa}. We generate events for the LHC at the CM energy $\sqrt{S} = 13$~TeV. We used the dynamic factorisation and renormalisation scale for the signal as well as the background events.

For the generation of parton-level events we apply minimum or maximum cuts on the transverse momentum $p_T$ and rapidities $\eta$ of light jets, $b$-jets, leptons, photons and missing transverse momentum ${\slashed {E}}_T$. Also, distance cuts between all possible final objects in the rapidity-azimuthal plane are applied, with the distance between two objects $i$ and $j$ defined as $\Delta R_{ij} = \sqrt{(\phi_i - \phi_j)^2 + (\eta_i - \eta_j)^2}$, where $\phi_i$ and $\eta_i$ are the azimuthal angle and rapidity of the object $i$, respectively.

The preliminary selection cuts are:
\begin{itemize}
  \item $p_T > 10$ and $|\eta| < 5$ for all $non$-$b$-jets, photons and leptons, and
  \item $\Delta R_{ij} > 0.4$ between all possible jets and leptons or photons.
\end{itemize}
After this, the {\scshape .lhe} files obtained through parton level events are showered with final state radiation (FSR) with {\texttt Pythia 8} where initial state radiation (ISR) and multiple interactions are switched off and fragmentation/hadronization is allowed.

We studied the following signal processes for the $2\ell + \slashed{E}_T$ final state.
\begin{description}[leftmargin=0pt,labelindent=0pt]
  \item[{\bf Sig1:}] $pp \to A_0 H_0$, with $A_0$ further decaying through the channel, $A_0 \to \ell^+ \ell^- H_0$.
  \item[{\bf Sig2:}] $pp \to H^+ H^-$. $H^\pm$ can further decay into at a pair 
leptons in following three ways:
  \begin{enumerate}[a)]
    \item Both the charged Higgs bosons decay as $H^+ (H^-) \to \ell^+ (\ell^-) \nu_\ell(\bar\nu_\ell) A_0$, followed by $A_0 \to \ell^+ \ell^- H_0$;
    \item Either one of the charged Higgs bosons decays as $H^+ (H^-) \to \ell^+ (\ell^-) \nu_\ell(\bar\nu_\ell) A_0$, followed by $A_0 \to \ell^+ \ell^- H_0$, whereas the other one decaying through $H^+ (H^-) \to \ell^+ (\ell^-) \nu_\ell(\bar\nu_\ell) H_0$.
    \item Both the charged Higgs bosons decay as $H^+ (H^-) \to \ell^+ (\ell^-) \nu_\ell(\bar\nu_\ell) H_0$.
  \end{enumerate}
  \item[{\bf Sig3:}] $pp \to H^\pm H_0$. $H^\pm$ can further decay into at a pair leptons as $H^+ (H^-) \to \ell^+ (\ell^-) \nu_\ell(\bar\nu_\ell) A_0$, followed by $A_0 \to \ell^+ \ell^- H_0$
\end{description}
\begin{table}[!ht]
 \centering
 \begin{tabular}{|c|c|c|c|c|c|c|c|c|}
  \hline
      & $m_{H_0}$
              & $m_{A_0}$
                      & $m_{H^+}$
                              & $m_{X}$
                                      & \multirow{2}*{$\lambda_L$}
                                               & \multirow{2}*{$\lambda_{HX}$}
                                                        & \multirow{2}*{$\lambda_{\Phi X}$}
                                                                 & \multirow{2}*{$\Omega_{H_0} {\rm h^2}$}
                                                                             \\
      & [GeV] & [GeV] & [GeV] & [GeV] &        &        &        &           \\ \hline\hline
  BP1 & $150$ & $160$ & $160.1$ & $205$ & $0.01$ & $0.01$ & $0.07$ & $0.01390$ \\ \hline
  IDM & $150$ & $160$ & $160.1$ & $205$ & $0.01$ &        &        & $0.00468$ \\ \hline
  BP2 & $100$ & $125$ & $125.1$ & $155$ & $0.01$ & $0.01$ & $0.05$ & $0.00901$ \\ \hline
  IDM & $100$ & $125$ & $125.1$ & $155$ & $0.01$ &        &        & $0.00197$ \\ \hline
 \end{tabular}
 \caption{Viable benchmark points used for collider study. }
 \label{tab:bp}
\end{table}
We neglected any other signal process as they are negligible in comparison to the above-mentioned ones for our benchmark points given in \cref{tab:bp}. We have chosen these benchmark points on the basis of the analysis of the relic density described in the previous sections. In \cref{tab:bp} we have also given a comparison between IDM and our model for the same benchmark points.

The major background at the LHC for the $2\ell + \slashed{E}_T$ final state processes are as follows
\begin{description}[leftmargin=0pt,labelindent=0pt]
  \item[{\bf Bkg1:}] $pp \to t \bar t$, followed by the top (anti-)quark decaying into the leptonic channel, $t (\bar t) \to \ell^+ (\ell^-) \nu_\ell (\bar \nu_\ell) b (\bar b)$.
  \item[{\bf Bkg2:}] $pp \to W^+ W^-$. $W^\pm$ further decays via leptonic channel as $W^+ (W^-) \to \ell^+ (\ell^-) \nu_\ell (\bar \nu_\ell)$.
  \item[{\bf Bkg3:}] $pp \to W^\pm Z (\gamma^*)$, followed by $W^+ (W^-) \to \ell^+ (\ell^-) \nu_\ell (\bar \nu_\ell)$, and $Z /\gamma^*$ decays into leptonic channel, $Z (\gamma^*) \to \ell^+ \ell^-$.
  \item[{\bf Bkg4:}] $pp \to Z Z (\gamma^*)$, followed by leptonic decays $Z \to \nu_\ell \bar \nu_\ell$ and $Z (\gamma^*) \to \ell^+ \ell^-$.
\end{description}
\begin{table}[!ht]
  \centering
  \begin{tabular}{|c|c|c|c|c|c|}
    \cline{1-3}\cline{5-6}
    \multirow{2}*{Processes}
               & \multicolumn{2}{c|}{Cross section [fb]}
                                 && Processes  & Cross section [fb] \\ \cline{2-3}\cline{5-6}
               & BP1    & BP2    && {\bf Bkg1} & $21269.0$  \\ \cline{1-3}\cline{5-6}
    {\bf Sig1} & $3.40$ & $9.83$ && {\bf Bkg2} & $3127.47$  \\ \cline{1-3}\cline{5-6}
    {\bf Sig2} & $1.99$ & $4.59$ && {\bf Bkg3} & $402.68$   \\ \cline{1-3}\cline{5-6}
    {\bf Sig3} & $1.26$ & $2.64$ && {\bf Bkg4} & $272.43$   \\ \cline{1-3}\cline{5-6}
  \end{tabular}
 \caption{Cross sections of the signals and backgrounds.}
 \label{tab:cs}
\end{table}
\cref{tab:cs} shows the cross sections of the signal processes for the above-mentioned benchmark points along with their backgrounds.

\begin{figure}[!ht]
  \centering
  \subfloat[]{\includegraphics[width=\linewidth]{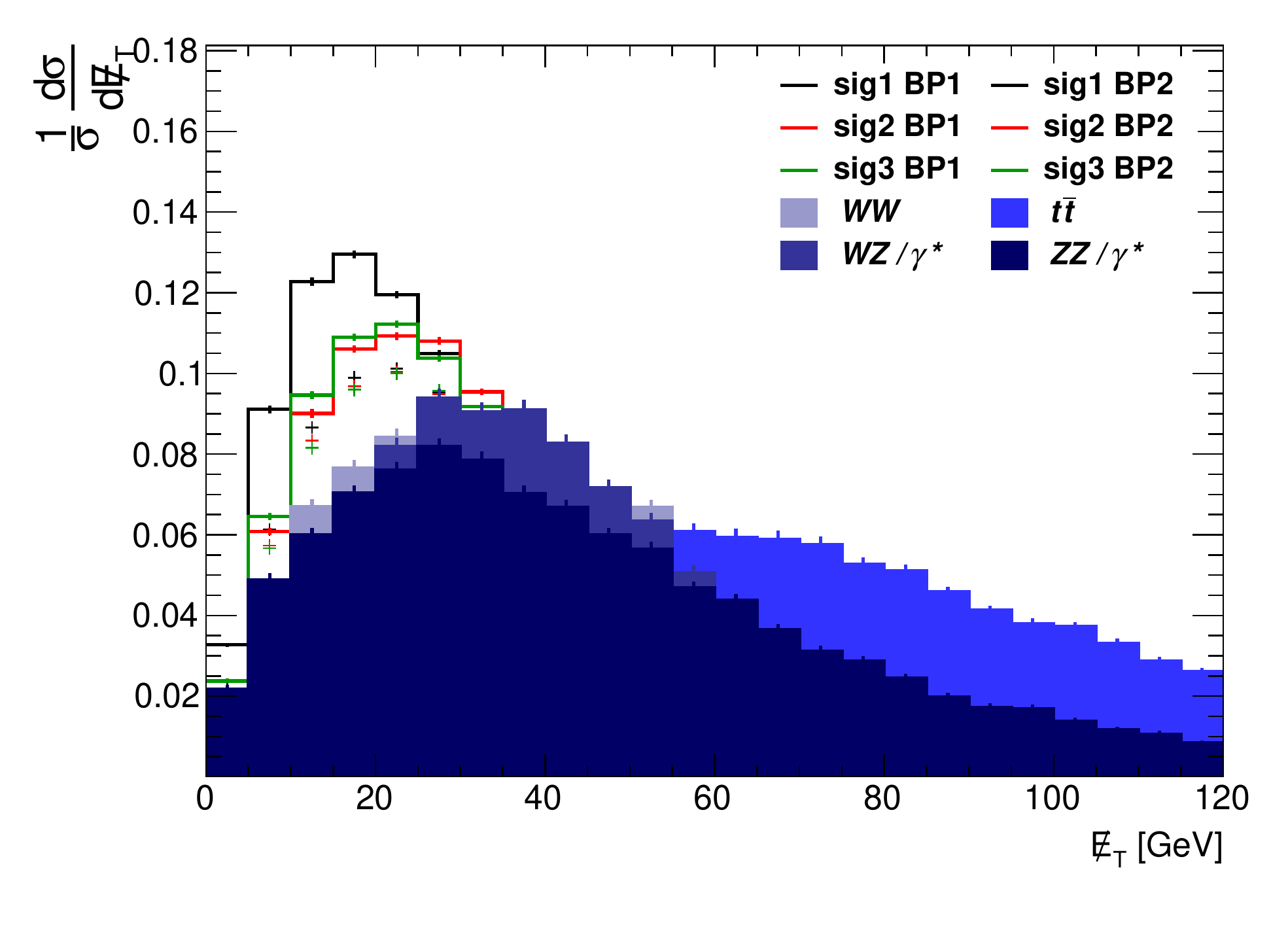}\label{fig:fig13a}}
  \\
  \subfloat[]{\includegraphics[width=\linewidth]{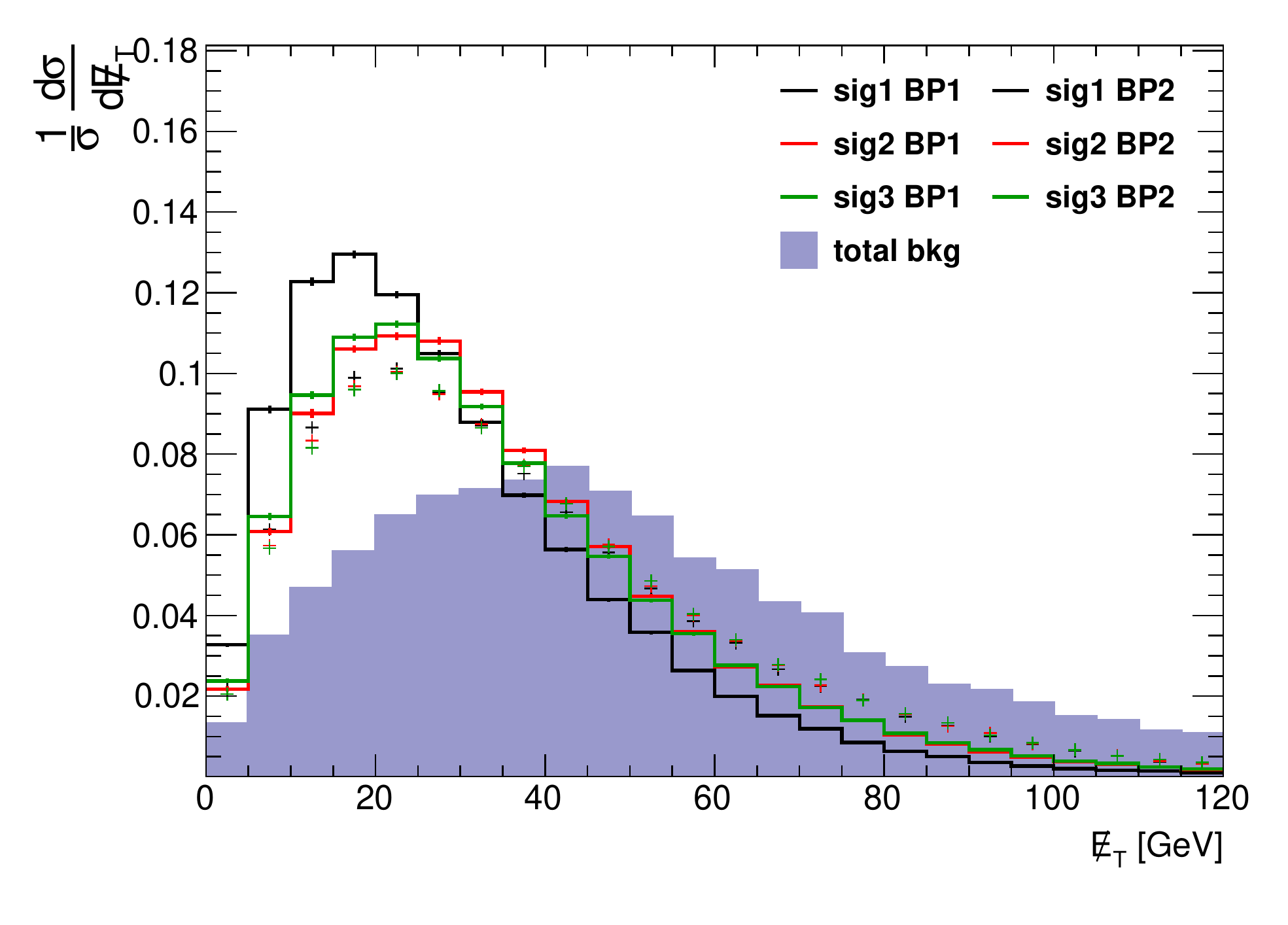}\label{fig:fig13b}}
  \caption{Distribution of $\slashed{E}_T$ for the benchmark points given in \cref{tab:bp}. The solid histograms belong to the BP2 and the dashed histograms (with only the peak value in each bin) show that for BP1. In the top panel, we show the effects of each background individually and in the bottom that of the total backgrounds.}
  \label{fig:fig13}
\end{figure}

As is obvious from \cref{tab:cs} that the cross section is greater in the low mass region than the higher one, we can na{\"i}vely assume that the lower mass region holds much more promise in the search of new BSM signal. In \cref{fig:fig13} we showed the $\slashed{E}_T$ distribution for our benchmark points which further strengthens our conviction. Please note the difference between the two plots of \cref{fig:fig13}. In \cref{fig:fig13a} we plotted the $\slashed{E}_T$ distribution with each individual background. Here we see that the signal is almost entirely overshadowed by the background which conforms with previous such studies. However \cref{fig:fig13b} shows that if we see the signal with the total normalised background we can clearly distinguish the signals from the backgrounds for a low mass region. Another important point in our study is that we refrain from a strong $\slashed{E}_T$ cut. Usually, the LHC searches for the dark matter was conducted in the context of SUSY theories. As a result, a strong $\slashed{E}_T$ cut $\sim 100$~GeV is applied to suppress the background which washes out the signal entirely in the low $\slashed{E}_T$ region. Here such a strong $\slashed{E}_T$ cut is not necessary for IDM alone, specifically for the benchmark points of our choice.

Finally, we can say that a large $\slashed E_T$ signal at the LHC where the dark matter can be produced along with other visible Standard Model particles, be it photons, jets or leptons, as in our case, can provide a mode of discovery of invisible sector. Although the measurement of $\slashed E_T$ relies heavily on the precise measurement of all the other visible particles in the collision, it is a powerful tool for observing dark matter. With the proposed upgrade in luminosity coming up, if we succeed in detecting dark matter at the LHC, it will further complement the searches from cosmic ray experiments and help us solve the riddle of the Universe.

%
\section{Conclusion}
\label{sec:conc}
In this work, we perform a detailed analysis of a multicomponent dark matter model. We consider a two-component dark matter model with inert Higgs doublet associated with a vector boson dark matter both odd under two different discrete symmetries. We explore the intermediate mass regime of inert Higgs doublet in the range $100-500$~GeV and observe that in presence of the other DM candidate it is possible to enhance the relic density of inert dark matter $H_0$ formidably when compared with the usual single component inert doublet model. We observe that the allowed mass ranges of inert doublet and vector dark matter which satisfies total DM relic abundance depend significantly on the coupling between inert Higgs doublet and vector dark matter $\lambda_{\Phi X}$. We scan over a range of mass for both $H_0$ and $X$ in the range $100~{\rm~GeV}\leq m_{H_0,X}\leq 500$~GeV for a specific value of $\lambda_{L}=\lambda_{HX}=0.01$ and mass splitting $\Delta m$. We show that there exists a large allowed parameter space of the model in $m_{H_0}$ vs $m_{X}$ plane for different $\lambda_{\Phi X}$ values satisfying the condition $m_{H_0}\leq m_{X}$. Direct detection limits from XENON1T further restricts the model parameter space by excluding regions below $m_{X}< 140$~GeV. However, this can be avoided by considering a smaller $\lambda_{HX}$ that satisfies direct detection limits for $m_{X}\sim 100$~GeV making the total range $100\leq m_{X}\leq 500$~GeV accessible. But this does not affect the phenomenology of multicomponent dark matter model. In a similar vein, we have shown that an allowed parameter space can be obtained in the above mentioned intermediate mass range which obeys the indirect detection bound from Fermi-LAT. Since IDM already satisfies relic bound in the region \(m_{H_0} \geq 550\)~GeV onwards due to co-annihilation channels which become more effective in this mass range, we have not shown the effect of our analysis in this mass range in the paper. But we checked the region for both DM masses up to 2 TeV, the features remain the same - only larger couplings are required in the higher mass range.

Apart from the phenomenology of the dark sector, we also explore the collider signature of the inert doublet dark matter. We observe a clear signal in the lower mass region of the new particles of the inert doublet model. However one should be careful in putting a hard $\slashed{E}_T$ cut. Otherwise, the signal will completely be washed away. The collider signatures of vector dark matter are not very significant and have not been pursued in this work. In conclusion, we show that the present model with inert doublet and a vector boson dark matter can successfully provide a multicomponent dark matter scenario.

\section*{Acknowledgements}
We acknowledge fruitful discussions with Poulose Poulose, Purusottam Ghosh and Basabendu Barman. RI thanks the SERB-DST, India for the research grant EMR/2015/000333. ADB acknowledges the support from the Department of Science and Technology, Government of India under the fellowship reference number PDF/2016/002148 (SERB National Post-Doctoral fellowship). SC would like to thank MHRD, Government of India for the research fellowship.

\bigskip

\appendix
\section{Vector DM annihilation cross sections}
\label{app:appA}
In this appendix, we give the analytical expressions of the most important conversion cross sections for vector DM.
  \begin{align}
   &\sigma_{XX \to H^+H^-}
   =
   \frac{1}{144 \pi s m_X^4}
   \frac{\sqrt{s-4 m_{H^\pm}^2}}{\sqrt{s-4 m_X^2}}
   \nonumber\\
   &\times
    \bigg[
    4 \frac{(\lambda_L v^2 + m_{H^\pm}^2 - m_{H_0}^2)^2 \lambda_{HX}^2}{(s - m_h^2)^2}
   \nonumber\\
   &
   - 4 \frac{(\lambda_L v^2 + m_{H^\pm}^2 - m_{H_0}^2) \lambda_{HX} \lambda_{\Phi X}}{(s-m_h^2)}
   + \lambda_{\Phi X}^2
   \bigg]
   \nonumber\\
   &\times
   \left(12 m_X^4-4 m_X^2 s+s^2\right)
  \end{align}
  \begin{align}
   &\sigma_{XX \to H_0H_0}
   =
   \frac{1}{288 \pi s m_X^4}
   \frac{\sqrt{s-4 m_{H_0}^2}}{\sqrt{s-4 m_X^2}}
   \nonumber\\
   &\times
    \bigg[
    \frac{\lambda_L^2 \lambda_{HX}^2 v^4}{(s - m_h^2)^2}
   - 2 \frac{\lambda_L \lambda_{HX} \lambda_{\Phi X} v^2}{(s-m_h^2)}
   + \lambda_{\Phi X}^2
   \bigg]
   \nonumber\\
   &\times
   \left(12 m_X^4-4 m_X^2 s+s^2\right)
  \end{align}
  \begin{align}
   &\sigma_{XX \to A_0A_0}
   =
   \frac{1}{288 \pi s m_X^4}
   \frac{\sqrt{s-4 m_{A_0}^2}}{\sqrt{s-4 m_X^2}}
   \nonumber\\
   &\times
    \bigg[
    4 \frac{(\lambda_L v^2 + m_{A_0}^2 - m_{H_0}^2)^2 \lambda_{HX}^2}{(s - m_h^2)^2}
   \nonumber\\
   &
   - 4 \frac{(\lambda_L v^2 + m_{A_0}^2 - m_{H_0}^2) \lambda_{HX} \lambda_{\Phi X}}{(s-m_h^2)}
   + \lambda_{\Phi X}^2
   \bigg]
   \nonumber\\
   &\times
   \left(12 m_X^4-4 m_X^2 s+s^2\right)
  \end{align}

\bibliography{i2HDM}
\bibliographystyle{apsrev4-1}

\end{document}